\documentclass[aps,floats]{revtex4-2}
\usepackage{amsmath,amssymb}
\usepackage{graphicx,epsfig}
\usepackage[greek,english]{babel}
\usepackage{bbold}

\begin{document}
\bibliographystyle {plain}

\pdfoutput=1
\def\oppropto{\mathop{\propto}} 
\def\opsimeq{\mathop{\simeq}}
\def\opoverderline{\mathop{\overline}}
\def\operarrow{\mathop{\longrightarrow}}
\def\opsim{\mathop{\sim}}

\def\opmin{\mathop{\min}} 
\def\opmax{\mathop{\max}} 
\def\oplim{\mathop{\lim}}

\title{  Large deviations and conditioning for chaotic non-invertible deterministic maps: \\
analysis via the forward deterministic dynamics and the backward stochastic dynamics } 


\author{C\'ecile Monthus}
\affiliation{Universit\'e Paris-Saclay, CNRS, CEA, Institut de Physique Th\'eorique, 91191 Gif-sur-Yvette, France}


\begin{abstract}
The large deviations properties of trajectory observables for chaotic non-invertible deterministic maps as studied recently by N. R. Smith, Phys. Rev. E 106, L042202 (2022) and by R. Gutierrez, A. Canella-Ortiz, C. Perez-Espigares, arXiv:2304.13754 are revisited in order to analyze in detail the similarities and the differences with the case of stochastic Markov chains. To be concrete, we focus on the simplest example displaying the two essential properties of local-stretching and global-folding, namely the doubling map  $ x_{t+1} = 2 x_t  [\text{mod} 1] $ on the real-space interval $x \in [0,1[$ that can be also analyzed via the decomposition $x= \sum_{l=1}^{+\infty} \frac{\sigma_l}{2^l} $ into binary coefficients $\sigma_l=0,1$. The large deviations properties of trajectory observables can be studied either via deformations of the forward deterministic dynamics or via deformations of the backward stochastic dynamics. Our main conclusions concerning the construction of the corresponding Doob canonical conditioned processes are: (i) non-trivial conditioned dynamics can be constructed only in the backward stochastic perspective where the reweighting of existing transitions is possible, and not in the forward deterministic perspective; (ii) the corresponding conditioned steady state is not smooth on the real-space interval $x \in [0,1[$ and can be better characterized in the binary space $\sigma_{l=1,2,..,+\infty}$. As a consequence, the backward stochastic dynamics in the binary space is also the most appropriate framework to analyze higher levels of large deviations, and we obtain the explicit large deviations at level 2 for the probability of the empirical density of long backward trajectories.

\end{abstract}

\maketitle

 \section{ Introduction }

 \subsection{ Large deviations for dynamical trajectories : deterministic chaotic systems versus Markov processes  }

The study of large deviations properties of deterministic dynamical systems has a long history (see the review
\cite{oono} and the PhD thesis \cite{tailleur_thesis} with references therein), in parallel with the general
theory of large deviations in statistical physics \cite{oono,ellis,review_touchette} and with the recent 
developments for stochastic processes (see the reviews with different scopes \cite{derrida-lecture,harris_Schu,searles,harris,mft,sollich_review,lazarescu_companion,lazarescu_generic,jack_review}, 
the PhD Theses \cite{fortelle_thesis,vivien_thesis,chetrite_thesis,wynants_thesis,chabane_thesis,duBuisson_thesis} 
 and the Habilitation Thesis \cite{chetrite_HDR}).
  For Markov processes, the large deviations properties of 
 time-local trajectory observables over a large time-window $[0,T]$
have been much studied recently via the appropriate deformations of the Markov generators
 \cite{peliti,derrida-lecture,sollich_review,lazarescu_companion,lazarescu_generic,jack_review,vivien_thesis,lecomte_chaotic,lecomte_thermo,lecomte_formalism,lecomte_glass,kristina1,kristina2,jack_ensemble,simon1,simon2,tailleur,simon3,Gunter1,Gunter2,Gunter3,Gunter4,chetrite_canonical,chetrite_conditioned,chetrite_optimal,chetrite_HDR,touchette_circle,touchette_langevin,touchette_occ,touchette_occupation,garrahan_lecture,Vivo,c_ring,c_detailed,chemical,derrida-conditioned,derrida-ring,bertin-conditioned,touchette-reflected,touchette-reflectedbis,c_lyapunov,previousquantum2.5doob,quantum2.5doob,quantum2.5dooblong,c_ruelle,lapolla,c_east,chabane,us_gyrator,duBuisson_gyrator,c_largedevpearson},
 and the construction of the corresponding canonical conditioned processes \cite{chetrite_conditioned,chetrite_optimal}.

 The same idea of deformed generators has been applied recently to 
 analyze the statistics of trajectory observables in various deterministic dynamical systems,
 either in continuous time \cite{tailleur_thesis,tailleur_Lyapunov,laffargue}
 with the development of population cloning algorithms,
 or in discrete time for chaotic non-invertible maps \cite{naftali,spain},
 that will be the focus of the present paper.
 Many analytical and numerical results with figures can be found in \cite{naftali,spain}
 for various trajectory observables
 of the most famous one-dimensional maps, namely the doubling map, the tent map,
 and the logistic map (see also \cite{naftali} for results on the two-dimensional cat map),
 with the following slightly different perspectives :
  \cite{naftali} focuses mainly on the dominant right-eigenvector eigenvalue equation of the deformed
 Frobenius-Perron operators 
 and on the Monte-Carlo algorithm 
 to construct the biased backward stochastic trajectories, 
 while \cite{spain} considers also the dominant left-eigenvectors of the deformed
 Frobenius-Perron operators and the construction 
 of the Doob conditioned dynamics making rare events typical. 
 
 
  \subsection{ Goals of the present work }
 
  The main goal of the present work is to revisit these two recent papers \cite{naftali,spain}
  in order to stress the similarities and the differences with the theory of large deviations
  at various levels for Markov processes, as summarized in Appendix \ref{app_markov}.
  In particular, we will discuss in detail the three following issues :
  
(1)  {\it What are the differences between the forward and the backward perspectives for chaotic non-invertible maps?}
  
 Chaotic non-invertible maps can be analyzed
either from the point of view of the forward deterministic dynamics
or from the point of view of the backward stochastic dynamics.
As a consequence, the appropriate deformations of these dynamics 
needed to analyze the statistics of trajectory observables
will have completely different qualitative properties, 
and it is thus important to understand the advantages and the drawbacks 
of the forward and the backward perspectives
In particular, we will stress that the construction of the Doob canonical conditioned processes, 
where the various existing trajectories are reweighted, is possible only within the backward 
stochastic perspective, and not within the forward deterministic perspective.

(2) {\it What properties are singular in real-space for chaotic non-invertible maps? }
 
 While for Markov processes, one is used to eigenvectors that are regular in real space,
 one needs to be aware that for chaotic non-invertible maps,
 various properties are singular in real-space, in particular some dominant eigenvectors of the deformed dynamics :
for the deformed forward dynamics, the right-eigenvectors are smooth in real space,
but the left-eigenvectors are singular in real space, as already found numerically 
on the example in Fig S1 of \cite{spain}; 
for the deformed backward dynamics, the left-eigenvectors are smooth  
but the right-eigenvectors are singular, so that their products corresponding to the Doob conditioned
steady states are always singular in real space. 
We will stress that all the observables that are singular in real-space
can be better analyzed via the symbolic dynamics, and 
explain how explicit expressions can be obtained in simple cases.

(3) {\it What higher levels of large deviations can be written explicitly for chaotic non-invertible maps ?} 
  
  As the large deviations of trajectory observables are not explicit in general, 
  since one is not able to solve eigenvalues equations 
  for arbitrary deformations of the generators,
  it is interesting to consider higher levels of large deviations
  in order to see what is the smallest level that can be written explicitly.
    Indeed for Markov processes (see the reminder in Appendix A), 
    a major progress has been the formulation of the so-called level 2.5,
 where one can write explicit rates functions 
 for the joint distribution of the empirical density and of the empirical flows,
  in particular for discrete-time Markov chains 
\cite{fortelle_thesis,fortelle_chain,review_touchette,c_largedevdisorder,c_reset,c_inference,c_microcanoEnsembles,c_diffReg},
for continuous-time Markov jump processes with discrete configuration space
\cite{fortelle_thesis,fortelle_jump,maes_canonical,maes_onandbeyond,wynants_thesis,chetrite_formal,BFG1,BFG2,chetrite_HDR,c_ring,c_interactions,c_open,barato_periodic,chetrite_periodic,c_reset,c_inference,c_LargeDevAbsorbing,c_microcanoEnsembles,c_susyboundarydriven,c_diffReg,c_inverse},
for diffusion processes in continuous space
\cite{wynants_thesis,maes_diffusion,chetrite_formal,engel,chetrite_HDR,c_lyapunov,c_inference,c_susyboundarydriven,c_diffReg,c_inverse}, 
and for jump-diffusion or jump-drift processes \cite{c_reset,c_runandtumble,c_jumpdrift,c_SkewDB}.
All the lower levels can be then obtained via the optimization
of the explicit rate function 2.5 in the presence of the appropriate constraints, 
but the solutions of these
constrained optimizations cannot be written explicitly in general.
As a consequence, the large deviation at level 2 for the distribution of the empirical density alone,
as well as the large deviations of trajectory observables that can be rewritten in terms of
the empirical density and of the empirical flows,
are not explicit in general. 

For chaotic non-invertible maps, the backward dynamics is stochastic, 
but remains nevertheless very special
since the corresponding forward dynamics is deterministic.
We will stress that this property induces level reductions in the large deviations of backward trajectories 
with respect to the case of Markov chains : 
the application of the explicit level 2.5 for Markov chains actually reduces to an explicit expression
for the level 2 concerning the distribution of the empirical density alone;
the general trajectory observables of Markov chains that involve both
the empirical density and the empirical flows can be rewritten in terms of the empirical density alone,
and thus reduce to the trajectory observables of the so-called Level 1, 
that can be obtained from the Level 2 via contraction.

 
  \subsection{ Organization of the present paper and main results}

In order to discuss more concretely the three issues summarized above, 
we will focus on the simplest chaotic non-invertible displaying the two essential properties of local-stretching and global-folding, namely the doubling map on the real-space interval $x \in [0,1[$.
Let us now describe the purpose of each section and mention the equation numbers where
the main results can be found.

$\bullet$ In section \ref{sec_forward}, the essential properties of the forward deterministic dynamics
associated to the doubling map are recalled, both in real space and in binary space.
The spectral properties governing the convergence towards the uniform steady state
actually involve a lot of subtleties   
(see \cite{gaspard,linas,Hu22} and references therein). 
For our present purposes, the most important property  
is that the non-dominant left-eigenvectors of the Frobenius-Perron
are singular in real space (Eqs \ref{lndeltasingular} and \ref{lndeltasingularfirst}),
as a consequence of the sensitivity with respect to initial conditions.

$\bullet$ In section \ref{sec_forwardk}, the appropriate deformations of this forward deterministic dynamics
 are analyzed to study large deviations of trajectory observables,
that actually reduce to the observables of the so-called Level 1 (Eq. \ref{observablemap})
as a consequence of the deterministic character of the forward dynamics.
We stress that the dominant left-eigenvector of the deformed dynamics is singular in real space,
already at the first order of perturbation theory that involves the non-dominant left eigenvectors
of the unperturbed dynamics (Eq. \ref{eigenvecper1leftdoubling}).
As a consequence, the Doob conditioned steady state is also singular in real space,
while the Doob conditional kernel coincides with the initial kernel (Eq. \ref{WDoobkdeter}),
since the Doob canonical conditioning can only change the weights of the possible transitions
but cannot produce new transitions. 

$\bullet$ In section \ref{sec_backward}, the backward stochastic dynamics
associated to the doubling map is described, both in real space and in binary space,
in order to stress the differences with respect to the forward deterministic dynamics discussed in section 
\ref{sec_forwardk}.

$\bullet$ In section \ref{sec_backward}, the large deviations of trajectory observables
are analyzed via the deformations of this backward stochastic dynamics.
We stress that for this deformed backward dynamics,
the dominant left-eigenvector is regular in real space,
but the dominant right-eigenvector is singular in real space,
so that their product corresponding to the Doob conditioned
steady state is again always singular in real space. 
However here the Doob conditional backward kernel 
is different from the initial backward kernel 
and corresponds to the appropriate reweighting of the two pre-images
(Eqs \ref{WDoobkbackdoubling} and \ref{Ckprobadoubling} in real space,
or Eqs \ref{WDoobkbackbinary} and \ref{probaBCbinary} in binary space).

$\bullet$ In section \ref{sec_omegax}, all these previous different perspectives are illustrated
on the simplest trajectory observable of the doubling map
already considered in the two recent papers \cite{naftali,spain}.
We first recall the explicit solution \cite{naftali}
for the dominant eigenvalue (Eq \ref{lambdadoublingx}) of the deformed forward dynamics 
with its corresponding right eigenvector that is smooth in real space (Eq. \ref{rightdoublingx})
and factorized in binary space (Eq. \ref{r0sigmakx}).
We write the corresponding Doob conditioned backward kernel in real-space (Eq. \ref{WDoobkbackdoublingx})
as well as in binary space (Eq. \ref{WdoublingsigmabackwardkxConditioned}).
The conditioned steady state is given by a simple explicit expression in binary space (Eq. \ref{PdoublingsigmabackwardkxConditioned}) while its translation in real space is singular.
We also give the explicit expression in binary space of the dominant left eigenvector
of the deformed forward dynamics (Eq. \ref{l0sigmakx}), whose translation in real space is singular
as already found numerically on Fig S1 of \cite{spain}.

$\bullet$ In section \ref{sec_level2}, we turn to the third issue (3) described in the previous subsection
concerning higher levels of large deviations beyond the level of trajectory observables.
We stress that the application of the explicit large deviations at level 2.5 for arbitrary Markov chains
(Eqs \ref{level2.5chain} and \ref{rate2.5chain})
to the specific case of the stochastic backward dynamics associated to the doubling map
produces explicit large deviations at level 2 for the empirical density alone (Eqs \ref{level2backwardsigma}
and \ref{rate2backwardsigma}).
This level reduction from the level 2.5 towards the level 2 can be understood
from the very special type of stochasticity of the backward dynamics  
that is associated to a forward deterministic dynamics.

$\bullet$ In \ref{sec_conclusion}, we summarize 
our conclusions concerning the doubling map 
and discuss the generalizations to other chaotic non-invertible maps or other chaotic deterministic dynamics.

$\bullet$ Three appendices contain complementary material with respect to the main text:
  Appendix \ref{app_markov} summarizes  the large deviations properties at various levels
  for Markov chains,
  while Appendix \ref{app_backward} recalls the properties of the backward Markov chain
  associated to a given forward Markov chain;
  Appendix \ref{app_fourier} contains the properties of the doubling map for the Fourier coefficients
  of the density on the real space interval $x \in [0,1[$.


\section{ Doubling-map deterministic dynamics in real-space and in binary-space}

\label{sec_forward}

In this section, some important properties of the Doubling-map deterministic dynamics
are summarized 
in order to stress the similarities and differences with Markov chains described in section \ref{sec_markovchain} of Appendix \ref{app_markov}.
 

\subsection{ Deterministic trajectories on the real-space interval $[0,1[$}

The doubling map on the interval $x \in [0,1[$
\begin{eqnarray}
x_{t+1} = f(x_t) \equiv 2 x_t  [\text{mod} 1] \equiv  
\begin{cases}
2 x_t \ \ \ \ \ \ \ \ \ \text{  for $0 \leq x_t < \frac{1}{2}$}  
 \\
(2x_t-1) \text{ for $\frac{1}{2} \leq x_t < 1$}
\end{cases}
\label{doubling}
\end{eqnarray}
is the simplest example of chaotic dynamics displaying the two essential properties of local-stretching and global-folding: 

(i) the uniform local-stretching of amplitude $f'(x)=2$
ensures the exponential divergence in time of the separation between two nearby trajectories,
and thus the sensitivity to initial conditions.

(ii) the global-folding corresponds to the non-invertibility of the doubling map,
where each value $x_{t+1} \in [0,1[$ has two pre-images 
\begin{eqnarray}
x_t^- && = \frac{x_{t+1}}{2} \in [0,\frac{1}{2}[
\nonumber \\
x_t^+ && = \left( \frac{1}{2}+\frac{x_{t+1}}{2}\right) \in [\frac{1}{2},1[
\label{twopreimages}
\end{eqnarray}
so that some information about the past is lost at each time-step.
As a consequence, when one tries to reconstruct the past trajectories backward in time, one obtains a stochastic dynamics
as will be described in section \ref{sec_backward}, while in the present section, we focus on the forward deterministic point of view.


\subsection{ Frobenius-Perron dynamics for the
probability density $\rho_t(  x)  $ on the real space interval $x \in [0,1[$ }

The Heaviside function $\theta(z)$ defined as  
\begin{eqnarray}
\theta(z) = \begin{cases}
1 \text{  for $z \geq 0$}  
 \\
0 \text{ for $z<0$}
\end{cases}
\label{heaviside}
\end{eqnarray}
is useful to write the forward deterministic kernel $w({\tilde x} \vert x) $
associated to the doubling map of Eq. \ref{doubling} 
\begin{eqnarray}
w({\tilde x} \vert x)  \equiv \delta({\tilde x}-f(x) )
&& =\delta \big({\tilde x} - 2x \big) \left[ \theta \left( x \right) - \theta \left( x - \frac{1}{2} \right)\right]
+\delta \big({\tilde x} - (2x-1) \big) \left[ \theta \left( x - \frac{1}{2}\right) - \theta \left( x - 1 \right)\right]
\nonumber \\
&& 
= \frac{ \delta(x- \frac{ {\tilde x}}{2} ) + \delta(x-\frac{ {\tilde x}+1}{2} ) }{2 } 
\label{Wdoubling}
\end{eqnarray}
that governs the Frobenius-Perron dynamics 
for the probability density $\rho_t(  .)  $
\begin{eqnarray}
  \rho_t({\tilde x}) = \int_0^1 dx w({\tilde x} \vert x) \rho_{t-1}(x)
= \frac{  \rho_{t-1}( \frac{ {\tilde x}}{2} ) +  \rho_{t-1}( \frac{ {\tilde x}+1}{2} ) }{2 }
\label{doublingFP}
\end{eqnarray}
which is the analog of the Markov chain evolution of Eq. \ref{markovchainc}.
The physical interpretation is that the new histogram $\rho_t(  .) $
represents the average of the two rescaled half-histograms of $\rho_{t-1}(.) $ on the two intervals $[0,\frac{1}{2}]$ and  
$[\frac{1}{2},1]$.
The iteration of Eq. \ref{doublingFP}
up to the initial density $\rho_{t=0}(.)$ at $t=0$
\begin{eqnarray}
  \rho_t( x) && = 
\frac{  \rho_{t-2}( \frac{ x}{4} ) +  \rho_{t-2}( \frac{ x+1}{4} ) + \rho_{t-2}( \frac{ x+2}{4} ) +  \rho_{t-2}( \frac{ x+3}{4} )  }{4 }
= ...
 =\frac{ 1}{2^t} \sum_{j=0}^{2^t-1} \rho_0 \left(  \frac{ x}{2^t} +\frac{ j}{2^t}  \right)
\label{doublingFPiter}
\end{eqnarray}
yields that the histogram $\rho_t(  .) $ at time $t$
represents the average of the $2^t$ rescaled histograms $\rho_0(.) $ on the $2^t$ intervals $[\frac{ j}{2^t},\frac{ j+1}{2^t}]$ labelled by $j=0,1,..,2^t-1$.
If the initial density $\rho_{t=0}$ is smooth,
one thus expect the convergence towards the uniform invariant density
\begin{eqnarray}
\rho_t( x) \opsimeq_{t \to + \infty} \rho_*(  x) =1
\label{perronFrobsteady}
\end{eqnarray}
that corresponds to the steady solution for the dynamics of Eq. \ref{doublingFP}.
As for Markov chains, one would like to characterize the convergence 
convergence towards this steady state 
via the spectral decomposition of the kernel analogous to Eq. \ref{Wspectral}.


\subsection{ Spectral decomposition when 
the initial density $\rho_{t=0}(  x)  $ is infinitely differentiable on the interval $x \in [0,1[$ }

Let us summarize the spectral properties of the kernel $w({\tilde x} \vert x) $
in the space of infinitely differentiable densities on the interval $x \in [0,1[$
(see \cite{gaspard,linas,Hu22} and references therein).
If the density $\rho_t(x)$ on the whole interval $x \in [0,1[$ can be reproduced by its Taylor expansion around the origin 
\begin{eqnarray}
\rho_t(x) = \sum_{p=0}^{+\infty} \frac{ \rho_t^{(p)}(0) }{p!} x^p
\label{taylor}
\end{eqnarray}
the dynamics can be analyzed via the action of the kernel $w({\tilde x} \vert x)$ on the powers $x^p$ for $p=1,2,..,+\infty$
\begin{eqnarray}
  \int_0^1 dx w({\tilde x} \vert x) x^p
= \frac{  ( \frac{ {\tilde x}}{2} )^p + ( \frac{ {\tilde x}+1}{2} )^p }{2 }
= \frac{ {\tilde x}^p}{2^p} + \sum_{q=0}^{p-1} {\tilde x}^q \frac{ \binom{p}{q}}{2^{p+1}}
\label{doublingpowerp}
\end{eqnarray}
Since it produces the same power ${\tilde x}^p$ and the smaller powers ${\tilde x}^q$ with $q=0,..,p-1$,
the diagonal coefficients directly give the eigenvalues $\lambda_{p \ne 0}$ of the kernel $w(. \vert .)$
\begin{eqnarray}
\lambda_p = 2^{-p} \ \ \ p=1,2,..,+\infty
\label{eigenvaluesdoubling}
\end{eqnarray}
The corresponding right eigenvectors satisfying Eq. \ref{eigenW}
\begin{eqnarray}
\lambda_p \ r_p(  x)  =  \frac{  r_p(  \frac{ x}{2}) +  r_p(  \frac{ x+1}{2}) }{2 }  
\label{eigenWmapdoublingr}
\end{eqnarray}
are the Bernoulli polynomials $r_p(x)=B_p(x) $ of degrees $p$ \cite{gaspard,linas} :
beyond $r_0(x)=B_0(x)=1=\rho_*(x)$ corresponding to the 
uniform steady state discussed in Eq. \ref{perronFrobsteady},
the first members read
\begin{eqnarray}
r_1(x)  && =x - \frac{1}{2} \equiv B_1(x)
\nonumber \\
r_2(x) && = x^2 -x + \frac{1}{6} \equiv B_2(x)
\nonumber \\
r_3(x)  && = x^3 - \frac{3}{2} x^2 + \frac{1}{2} x \equiv B_3(x)
\label{bernouilli}
\end{eqnarray}
while their Fourier definition is given in Eq. \ref{bernoullifourier} for arbitrary $p$.
Beyond $l_0(x)=1$, the left eigenvectors satisfying Eq. \ref{eigenW} with 
the eigenvalues $\lambda_p=2^{-p}$ of Eq. \ref{eigenvaluesdoubling} with $p=1,2,..$
\begin{eqnarray}
0 \leq x < \frac{1}{2} : \ \ \ \ \lambda_p  l_p(  x) && = l_p( 2x ) 
\nonumber \\
\frac{1}{2} \leq x < 1 : \ \ \ \ \lambda_p  l_p(  x) && = l_p( 2x-1 ) 
\label{eigenWmapdoublingleft}
\end{eqnarray}
are however singular distributions $D_p(x) $  
that read in terms of the delta functions and their  
derivatives of arbitrary orders at the two boundaries $x=0$ and $x=1$ \cite{gaspard,linas}
\begin{eqnarray}
  l_p(  x)  = \frac{ (-1)^{p-1} }{ p!} \left[ \delta^{(p-1)}(x-1) - \delta^{(p-1)}(x)\right] \equiv D_p(x)
\label{lndeltasingular}
\end{eqnarray}
with the first members
\begin{eqnarray}
l_1(x) && =   \delta(x-1) - \delta(x) \equiv D_1(x)
\nonumber \\
l_2(x) && =  - \frac{ 1 }{ 2} \left[ \delta'(x-1) - \delta'(x)\right]  \equiv D_2(x)
\nonumber \\
l_3(x) && =  \frac{ 1 }{ 6} \left[ \delta''(x-1) - \delta''(x)\right]  \equiv D_3(x)
\label{lndeltasingularfirst}
\end{eqnarray}
The scalar product of the distribution $D_p(x)$ for $p>0$ with another function $g(x)$ 
can be evaluated
using $(p-1)$ integrations by parts 
\begin{eqnarray}
\langle D_p \vert g \rangle =  \int_0^1 dx D_p(x) g(x) 
&& =  \int_0^1 dx g(x) \frac{ (-1)^{p-1} }{ p!} \frac{d^{(p-1)}}{dx^{p-1}} \left[ \delta(x-1) - \delta(x)\right] 
 =  \int_0^1 dx \left[ \delta(x-1) - \delta(x)\right]  \frac{ 1 }{ p!} \frac{d^{(p-1)} g(x)}{dx^{p-1}} 
\nonumber \\
&&  =  \frac{ g^{(p-1)}(1) - g^{(p-1)}(0) }{ p!} 
\label{integparts}
\end{eqnarray}

In particular, the spectral decomposition of Eq. \ref{markovspectral} 
\begin{eqnarray}
\rho_t({\tilde x}) &&   =1 +  \sum_{p=1}^{+\infty} 2^{-pt}  B_p( {\tilde x}) \int_0^1 dx_0 D_p(x_0) \rho_0(x_0)
\nonumber \\
&& =1 +  \sum_{p=1}^{+\infty} 2^{-pt}  B_p( {\tilde x}) \left[\frac{ \rho_0^{(p-1)}(1) - \rho_0^{(p-1)}(0) }{ p!}  \right]
\label{perronFrobspectraldoubling}
\end{eqnarray}
that corresponds to the Euler-Maclaurin formula for $t=0$ \cite{gaspard},
can be used 
only for if the initial density $\rho_0(x)$ is infinitely differentiable.
Note that when the initial density $\rho_0(x)$ is not infinitely differentiable,
the discussion of the spectral properties is much more involved
and depends on the precise space of densities
 that one wishes to consider (see the very detailed discussion in \cite{linas}).
Besides the analysis of the dynamics on the real space interval $[0,1[$ discussed above, 
it is useful to consider other formulations (see the very detailed discussion in \cite{linas}),
in particular in Fourier space as recalled in Appendix \ref{app_fourier}
and in the binary-space as recalled in the next subsection.


\subsection{ Properties of the dynamics in terms of the binary coefficients $\sigma_l=0,1$ parametrizing $x=\displaystyle \sum_{l=1}^{+\infty} \frac{\sigma_l}{2^l} $  }

When $x \in [0,1[$ is represented by its binary coefficients $\sigma_l=0,1$ with $l=1,2,..+\infty$
\begin{eqnarray}
x  = \sum_{l=1}^{+\infty} \frac{\sigma_l}{2^l} =  \frac{\sigma_1}{2}+ \frac{\sigma_2}{4}+ \frac{\sigma_3}{8}+...
\label{binary}
\end{eqnarray}
the dynamics of the doubling map of Eq. \ref{doubling}
\begin{eqnarray}
{\tilde x } \equiv \sum_{l=1}^{+\infty} \frac{{\tilde \sigma}_l}{2^l} 
= 2 x  [\text{mod} 1] = \sum_{l=1}^{+\infty} \frac{\sigma_{l+1} }{2^l} 
\label{xbinaryimage}
\end{eqnarray}
translates into the shift dynamics for the binary coefficients
\begin{eqnarray}
{\tilde \sigma }_l=  \sigma_{l+1} \ \ \text{ for } l=1,2,..
\label{binaryimage}
\end{eqnarray}
while the first coefficient $\sigma_1 $ disappears.

The real-space kernel $w(. \vert .) $ of Eq. \ref{Wdoubling} translates for the binary coefficients
into the kernel
\begin{eqnarray}
W( {\tilde \sigma }_. \vert \sigma_.)  \equiv \prod_{l=1}^{+\infty} \delta_{{\tilde \sigma }_l,\sigma_{l+1}}
\label{Wdoublingsigma}
\end{eqnarray}
The normalization of this kernel over ${\tilde \sigma }_. $ corresponds to 
the left eigenvector unity for any configuration ${\tilde \sigma }_.  $
\begin{eqnarray}
L_0( {\tilde \sigma }_. )  = \prod_{l=1}^{+\infty} \left( \delta_{{\tilde \sigma }_l,0} + \delta_{{\tilde \sigma }_l,1}\right)  
\label{Wdoublingsigmaleft}
\end{eqnarray}

The correspondence between the probability density $\rho_t(x)$ on the interval $[0,1[$ 
and the probability $P_t(   \sigma_.) $ of the binary variables reads
\begin{eqnarray}
\rho_t(x)   
= \sum_{\sigma_.}  P_t(  \sigma_.)  \delta \left( x-\displaystyle \sum_{l=1}^{+\infty} \frac{\sigma_l}{2^l}\right)
\label{rhoxPsigma}
\end{eqnarray}

The steady uniform density $\rho_*(  x) =1 $ of Eq. \ref{perronFrobsteady}
translates into the steady uniform probability for the binary variables
$\sigma_l=0,1$ drwn with probabilities $\left(\frac{1}{2}, \frac{1}{2}\right)$
\begin{eqnarray}
P_*(  \sigma _. )  = \prod_{l=1}^{+\infty} \left( \frac{\delta_{ \sigma _l,0} +  \delta_{\sigma _l,1} }{2} \right)  
= R_0(  \sigma _. )
\label{Wdoublingsigmaright}
\end{eqnarray}
that is the right eigenvector $R_0(  \sigma _. )  $ of the kernel of Eq. \ref{Wdoublingsigma}
associated to the highest eigenvalue $\lambda_0=1$.

The Frobenius-Perron dynamics for the probability $P_t(.) $ of the binary variables  
\begin{eqnarray}
P_t(   {\tilde \sigma }_1, {\tilde \sigma }_2,...)  
= \sum_{\sigma_.} w( {\tilde \sigma }_. \vert \sigma_.) P_{t-1}(  \sigma_.)  
= \sum_{\sigma_1=\pm 1} P_{t-1}(  \sigma_1, {\tilde \sigma }_1, {\tilde \sigma }_2,...)  
\label{FPsigma}
\end{eqnarray}
means that one just integrates the previous probability $P_{t-1}(.)$ over its first variable $\sigma_1$.
The iteration up to the initial probability $P_{t=0}(.)$ at time $t=0$ 
\begin{eqnarray}
P_t(   {\tilde \sigma }_1, {\tilde \sigma }_2,...)  
=\sum_{\sigma_1=\pm 1; \sigma_2=\pm 1;...;\sigma_t=\pm 1}   P_0( \sigma_1,\sigma_2, ... ,\sigma_t,{\tilde \sigma }_1, {\tilde \sigma }_2,...)
\label{FPsigmaiter}
\end{eqnarray}
means that one just integrates the initial probability $P_{t=0}(.)$ over its first $t$ variables $(\sigma_1; \sigma_2;...;\sigma_t)$.
This formulation shows even more clearly the issues of
the convergence towards the uniform distribution of Eq. \ref{Wdoublingsigmaright} for large $t$
depending on the initial distribution $P_{t=0}(.)$ :
the convergence found in Eq. \ref{perronFrobsteady} in real space for initial densities that 
were sufficiently smooth
means for the binary variables that the coefficients $\sigma_l$ for large enough $l$
should be independent and take the values $(0,1)$ with the equal probabilities $\left(\frac{1}{2}, \frac{1}{2}\right)$.


\subsection{ Discussion : similarities and differences with Markov chains }

In summary, if the initial density $\rho_{t=0}(.)$ is infinitely differentiable on the real-space interval $x \in [0,1[$,
the convergence towards the uniform steady state via the spectral decomposition of Eq. \ref{perronFrobspectraldoubling}
is similar to the analog property of Eq. \ref{markovspectral} for Markov chains.
However, the fact that the left eigenvectors of Eq. \ref{lndeltasingular} are singular distributions
that reflects the sensitivity with respect to initial conditions
will have very important consequences in the study of large deviations for trajectory observables
via appropriate deformations of the kernel, as discussed in the next section.

\section{ Large deviations of trajectory observables via the deformed dynamics }

\label{sec_forwardk}

In this section, the analysis of large deviations of trajectory observables for chaotic non-invertible maps
  described in
  the two recent papers \cite{naftali,spain}
  is revisited on the concrete example of the doubling map 
  in order to analyze in detail the similarities and differences with the case of stochastic Markov chains
  summarized in subsection \ref{subsec_trajviadeform} of Appendix \ref{app_markov}.

The first important difference for any deterministic dynamics $x_{t+1}=f(x_t) $
is that the general trajectory observable of Eq. \ref{observablerho2}
which is constructed from the elementary transitions $x_t \to x_{t+1}$ during the long trajectory 
\begin{eqnarray}
{\cal O}^{Traj}\left[ x_0,x_1,...,x_T \right]  
&& =   \frac{1}{T  } \sum_{t=0}^{T-1} O ( x_{t+1},  x_t)
=  \frac{1}{T  } \sum_{t=0}^{T-1} O ( f(x_t),  x_t) 
\nonumber \\
&& \equiv  \frac{1}{T  } \sum_{t=0}^{T-1} \omega (  x_t)  
\ \ \ {\rm with } \ \ \omega (  x_t)=O ( f(x_t),  x_t) 
\label{observablemap}
\end{eqnarray}
reduces to an observable  
that involves only the function $\omega (  x_t)=\Omega ( f(x_t),  x_t)  $ of the single position $x_t$.


\subsection{  Dynamics of the generating function $Z_T^{[k]} (x_T \vert x_0)$ governed by the $k$-deformed kernel $w^{[k]}({\tilde x} \vert x)$  }

As recalled in details in subsection \ref{subsec_trajviadeform} of Appendix A, the standard method to analyze the large deviations of Eq. \ref{ldtraj}
for trajectory observables is based on the generating function $Z_T^{[k]} (x_T \vert x_0) $ of Eq. \ref{genektraj}
for the trajectory observable of Eq. \ref{observablemap}
\begin{eqnarray}
Z_T^{[k]} (x_T \vert x_0) \equiv  \langle e^{ \displaystyle k T {\cal O}^{Traj} } \rangle
&& = \int dx_1 \int dx_2 ... \int dx_{T-1} 
 \prod_{t=0}^{T-1} \left[ w(x_{t+1},x_t ) e^{ k \omega (  x_t)} \right]
 \nonumber \\
 && \equiv  \int dx_1 \int dx_2 ... \int dx_{T-1}
  \prod_{t=0}^{T-1} w^{[k]}(x_{t+1},x_t ) \equiv \langle x_T \vert  \left(w^{[k]}\right)^T \vert x_0\rangle 
\label{genektrajmap}
\end{eqnarray}
whose dynamics can be analyzed via
the $k$-deformation of Eq. \ref{Wkdeformed} 
that reads for the doubling-map kernel of Eq. \ref{Wdoubling}
\begin{eqnarray}
 w^{[k]}({\tilde x} \vert x) \equiv \delta({\tilde x}-f(x)) e^{k \omega(x)}
&& = 
\delta \big({\tilde x} - 2x \big) e^{k \omega(x)} 
\left[ \theta \left( x \right) - \theta \left( x - \frac{1}{2} \right)\right]
+\delta \big({\tilde x} - (2x-1) \big) e^{k \omega(x)} 
\left[ \theta \left( x - \frac{1}{2}\right) - \theta \left( x - 1 \right)\right]
\nonumber \\
&& 
= \frac{ \delta(x- \frac{ {\tilde x}}{2} ) + \delta(x-\frac{ {\tilde x}+1}{2} ) }{2 } e^{k \omega(x)}\left[ \theta \left({\tilde x} \right) - \theta \left({\tilde x} - 1 \right)\right]
\label{Wkdeformedmapdoubling}
\end{eqnarray}

The dynamics for the generating function
\begin{eqnarray}
Z^{[k]}_{T}({\tilde x})   = \int_0^1 dx w^{[k]}({\tilde x} \vert x) Z^{[k]}_{T-1}(x)
= \frac{ Z^{[k]}_{T-1}(\frac{ {\tilde x}}{2}) e^{k \omega(\frac{ {\tilde x}}{2})}
+ Z^{[k]}_{T-1}(\frac{ {\tilde x}}{2}) e^{k \omega(\frac{ {\tilde x}+1}{2} )}}{2}
\label{Wkdeformedmapdoublingiter}
\end{eqnarray}
corresponds to the deformation of the Frobenius-Perron dynamics of Eq. \ref{doublingFP}
for the probability density $\rho_t(  x)  $.


\subsection{ Eigenvalue problem governing the generating function $Z_T^{[k]} (x_T \vert x_0)$ for large time $T \to + \infty$  }

As recalled around Eq. \ref{Wktiltprop} for Markov chains,
 the generating function for large time $T$
is governed
by
the highest eigenvalue $\lambda_0^{[k]}$ associated to the positive right eigenvector 
$r_0^{[k]}( {\tilde x}) \geq 0 $ and to the positive left eigenvector $l_0^{[k]}( {\tilde x}) \geq 0 $
satisfying Eq. \ref{eigenWk}
that reads for the deformed doubling-map kernel of Eq. \ref{Wkdeformedmapdoubling}
\begin{eqnarray}
\lambda_0^{[k]}  r_0^{[k]}( {\tilde x}) && =  \int_0^1d  x  \ w^{[k]}({\tilde x} \vert x)  r_0^{[k]}(  x) 
=   \frac{  e^{ k \omega (\frac{ {\tilde x}}{2})} r_0^{[k]}(  \frac{ {\tilde x}}{2})
 +  e^{ k \omega (\frac{ {\tilde x}+1}{2})} r_0^{[k]}(  \frac{ {\tilde x}+1}{2})}{2 }
\nonumber \\
\lambda_0^{[k]}  l_0^{[k]}(  x) &&  =  \int_0^1d{\tilde x}  l_0^{[k]}( {\tilde x}) w^{[k]}({\tilde x} \vert x) 
\nonumber \\
&&  = l_0^{[k]}( 2x ) e^{ k \omega ( x)} \left[ \theta \left( x \right) - \theta \left( x - \frac{1}{2} \right)\right] 
+ l_0^{[k]}(2x-1) e^{ k \omega ( x)}\left[ \theta \left( x - \frac{1}{2}\right) - \theta \left( x - 1 \right)\right]
\label{eigenWkmapdoubling}
\end{eqnarray}
However, as already discussed for the undeformed case $k=0$ around Eq. \ref{perronFrobsteady},
the convergence of $\rho_t(x) $ towards the uniform distribution $\rho_*(x)=1 $ requires the smoothness of the initial condition $\rho_{t=0}(x_0)$.
Similarly, the convergence of the dynamics Eq. \ref{Wkdeformedmapdoublingiter} towards
\begin{eqnarray}
Z^{[k]}_{T}({\tilde x})  \opsimeq_{T \to + \infty} 
 [\lambda_0^{[k]}]^T r_0^{[k]}(x_T) \int dx_0  l_0^{[k]}(x_0) Z^{[k]}_{0}(x_0)
\label{Wkdeformedmapdoublingiterlarge}
\end{eqnarray}
requires the smoothness of the initial condition $Z^{[k]}_{T=0}(x_0)=\rho_{t=0}(x_0) $.

Then if one constructs the probability of the end-point analog to Eq. \ref{probaEndk},
it will converge for large time $T \to + \infty$
\begin{eqnarray}
P^{[k]End}_T(x ) 
\equiv \frac{Z_T^{[k]} (x }{ \int dy Z_T^{[k]} (y ) }
\opsimeq_{T \to +\infty}
\frac{r_0^{[k]}(x)}{ \int dy r_0^{[k]}(y)} \equiv P^{[k]End}_*(x)
 \label{probaEndkZ}
\end{eqnarray}
towards the distribution $P^{[k]End}_*(x) $ determined by the smooth right eigenvector $r_0^{[k]}(.) $.

However, if one tries to construct 
the conditioned kernel of Eq. \ref{WDoobk}
\begin{eqnarray}
w^{C[k]}({\tilde x} \vert x) 
&& \equiv  \frac{1}{\lambda_0^{[k]}} \  l_0^{[k]}({\tilde x}) w^{[k]}({\tilde x} \vert x)   \frac{1}{ l_0^{[k]}(x)  }
 = \frac{l_0^{[k]}({\tilde x}) w({\tilde x} \vert x) e^{ k \Omega ({\tilde x},x)}}{\int_0^1d  z  l_0^{[k]}(  z) w(z \vert x) e^{ k \Omega ( z,x)} }
 \nonumber \\
&& = \frac{l_0^{[k]}({\tilde x}) \delta({\tilde x} -f(x) ) e^{ k \omega ( x)}}
{\int_0^1d  z  l_0^{[k]}(  z) \delta(z -f(x)) e^{ k \omega ( x)} }
=\frac{l_0^{[k]}( f(x)) \delta({\tilde x} -f(x) ) e^{ k \omega ( x)}}
{  l_0^{[k]}(  f(x))  e^{ k \omega ( x)} }
=  \delta({\tilde x} -f(x) ) =  w({\tilde x} \vert x)   
  \label{WDoobkdeter}
\end{eqnarray}
one obtains that it coincides with the initial deterministic kernel $w({\tilde x} \vert x) = \delta({\tilde x} -f(x) ) $,
as expected since the Doob canonical conditioning can only change the weights of the possible transitions
as recalled after Eq. \ref{WDoobk},
but cannot produce new transitions.

On the other hand, the conditioned density given by Eq. \ref{ProbaDoobksteady}
in terms of the deformed eigenvectors of Eq. \ref{eigenWkmapdoubling}
\begin{eqnarray}
 \rho^{C[k]}_*( x)  \equiv  l_0^{[k]}(x)  r_0^{[k]}(x) 
  \label{ProbaDoobksteadydef}
\end{eqnarray}
should be different from the undeformed uniform steady density $\rho_*(x)=1$.
In order to clarify what is really going on,
it is thus useful to consider the perturbation theory in the deformation parameter $k$.


\subsection{ Perturbation theory in $k$ for any trajectory observable   }

\label{subsec_perturbation}

The perturbation theory in the deformation parameter $k$ 
is recalled for Markov chains in subsection \ref{subsec_permarkov} of Appendix \ref{app_markov}.
For the doubling map where the invariant density is uniform $\rho_*(x)=1$,
the first-order correction of Eq. \ref{lambdaper1}
reads for $ \omega (x)=\Omega(f(x),x)$ of Eq. \ref{observablemap}
using the normalization over ${\tilde x}$ of the kernel $w({\tilde x} \vert x) $
\begin{eqnarray}
  \lambda^{(1)}  = \int_0^1dx \left[\int_0^1d {\tilde x}  w({\tilde x} \vert x) \right] \omega (x)   \rho_* (x) = \int_0^1dx  \omega (x) 
\label{lambdaper1doubling}
\end{eqnarray}

The second-order correction $\lambda^{(2)} $ of Eq. \ref{lambdaper2}
reads using the other eigenvalues $\lambda_p = 2^{-p} $ of Eq. \ref{eigenvaluesdoubling} for $p=1,2,..,+\infty$,
with their right eigenvectors given by the Bernoulli polynomials $r_p(.)=B_p(.) $ of Eq. \ref{bernouilli}
and their left eigenvectors given by the singular distributions $l_p(.)=D_p(.)$ of Eq. \ref{lndeltasingular}
\begin{eqnarray}
 \lambda^{(2)}  && =     \int_0^1dx  \frac{ \omega^2 (x)} {2}  \rho_* (x)
   +  \sum_{p=1}^{+\infty} \frac{ \lambda_p }{1-\lambda_p}
   \left( \int_0^1dx_2   \omega(x_2)   r_p (x_2)  \right)
   \left( \int_0^1dx_1   l_p (x_1)  \omega(x_1)  \rho_* (x_1) \right)
   \nonumber \\
   && =   \int_0^1dx  \frac{ \omega^2 (x)} {2}  
   +  \sum_{p=1}^{+\infty} \frac{ 1 }{ 2^p-1}
   \left( \int_0^1dx_2   \omega(x_2)   B_p (x_2)  \right)
   \left( \int_0^1dx_1   D_p (x_1)  \omega(x_1)   \right)
\label{lambdaper2doubling}
\end{eqnarray}
where the last scalar product reads using Eq. \ref{integparts}
\begin{eqnarray}
\int_0^1dx_1   D_p (x_1)  \omega(x_1)  
 =  \frac{ \omega^{(p-1)}(1) - \omega^{(p-1)}(0) }{ p!} 
\label{integpartsomega}
\end{eqnarray}

The first-order perturbation theory of Eq. \ref{eigenvecper1} for the right eigenvector 
yields that the correction $ r^{(1)}_0 ({\tilde x}) $
with respect to the uniform invariant density $\rho_*(x)=1$
 can be decomposed onto the Bernoulli polynomials $B_p(.)$ of Eq. \ref{bernouilli}
\begin{eqnarray}
   r^{(1)}_0 ({\tilde x}) &&  =   \sum_{p=1}^{+\infty}  B_p ({\tilde x}) 
   \left[  \frac{ 1  }{2^p-1}  \int_0^1dx    D_p (x)   \omega (x)    \right]
   \nonumber \\
&& =    \sum_{p=1}^{+\infty}  B_p ({\tilde x}) 
   \left[  \frac{ \omega^{(p-1)}(1) - \omega^{(p-1)}(0) }{ (2^p-1)  p!}     \right]
\label{eigenvecper1rightdoubling}
\end{eqnarray}
 while the coefficients using Eq. \ref{integpartsomega}
 involve derivatives of arbitrary order of $\omega(.)$ at $x=0$ and $x=1$.

The first-order perturbation theory of Eq. \ref{eigenvecper1} for the left eigenvector 
yields that the correction $ l^{(1)}_0 ({\tilde x}) $
with respect to the trivial eigenvector $l_0(x)=1$
 can be decomposed onto the singular distributions $D_p(.)$ of Eq. \ref{lndeltasingular}
\begin{eqnarray}
  l^{(1)}_0 ({\tilde x})    = \sum_{p=1}^{+\infty}  D_p ({\tilde x}) 
  \left[ \frac{ 1 }{1- 2^{-p}}  \int_0^1dx  \omega(x)  B_p (x) \right]
\label{eigenvecper1leftdoubling}
\end{eqnarray}
where the coefficients can be computed as integrals of $\omega(x)$ with the Bernoulli polynomials $B_p(x)$
of Eq. \ref{bernouilli}.

The first-order correction of Eq. \ref{rhosteadyDoobper1}
for the conditioned density of Eq. \ref{rhosteadyDoob}
\begin{eqnarray}
 \rho^{C(1)}_*({\tilde x})  =  r_0^{(1)}({\tilde x})+l_0^{(1)}({\tilde x}) 
 =   \sum_{p=1}^{+\infty}  B_p ({\tilde x}) 
   \left[  \frac{ \omega^{(p-1)}(1) - \omega^{(p-1)}(0) }{ (2^p-1)  p!}     \right]
   +  \sum_{p=1}^{+\infty}  D_p ({\tilde x}) 
  \left[ \frac{ 1 }{1- 2^{-p}}  \int_0^1dx  \omega(x)  B_p (x) \right]
  \label{rhosteadyDoobper1doubling}
\end{eqnarray}
then contains both the regular contributions given by the Bernoulli polynomials $B_p({\tilde x})$ of Eq. \ref{bernouilli}
and the singular contributions given by the singular distributions $D_p(.)$ of Eq. \ref{lndeltasingular}.


\subsection{ Discussion : similarities and differences with Markov chains} 

The method to analyze large deviations of trajectory observables via 
the appropriate deformations of the dynamics is similar to the same approach for Markov chains,
but there are two essential differences:

(i) the conditioned kernel $w^{C[k]}({\tilde x} \vert x) $ constructed via Eq. \ref{WDoobkdeter}
coincides with the initial deterministic kernel $
w({\tilde x} \vert x)=  \delta({\tilde x} -f(x) ) $,
 since the Doob canonical conditioning can only change the weights of the possible transitions
but not create new transitions.

(ii) the first-order theory in $k$ indicates that 
both the deformed left eigenvector (Eq. \ref{eigenvecper1leftdoubling})
 and the conditioned density (Eq. \ref{rhosteadyDoobper1doubling})
 become very singular in real space as soon as $k \ne 0$.

 In order to overcome the first difficulty (i), 
 the natural idea is to replace the forward deterministic dynamics
 by the backward stochastic dynamics described in the next section.
 The singular character (ii) of various observables in real-space for $k \ne 0$
 suggests that it will be useful to analyze this backward stochastic dynamics
not only in real-space but also in the binary space where its deformations will be easier to characterize.

\section{ Stochastic backward dynamics in real-space and in binary-space }

\label{sec_backward}
 
 In this section, we describe the backward stochastic dynamics associated to the doubling map
 that is obviously closer to the Markov chains described in Appendix \ref{app_markov}
 than the deterministic forward dynamics considered in the previous sections.


\subsection{ Stochastic backward dynamics governed by the kernel $w_B(x \vert  {\tilde x}) $  }

As recalled in Appendix \ref{app_backward} for Markov chains, 
the trajectories can be alternatively constructed backward in time
via the appropriate backward kernel.
For chaotic non-invertible maps, the advantages of this backward perspective
have been already stressed in \cite{naftali}.
For the doubling map, 
the backward kernel $w_B(x \vert  {\tilde x}) $ of Eq. \ref{Wbackward}
associated to the forward deterministic kernel of Eq. \ref{Wdoubling}
and to its uniform steady state $\rho_*(x )=1 $ of Eq. \ref{perronFrobsteady}
reads
\begin{eqnarray}
w_B(x \vert  {\tilde x}) \equiv  w({\tilde x} \vert x) 
&& =\delta \big({\tilde x} - 2x \big) \left[ \theta \left( x \right) - \theta \left( x - \frac{1}{2} \right)\right]
+\delta \big({\tilde x} - (2x-1) \big) \left[ \theta \left( x - \frac{1}{2}\right) - \theta \left( x - 1 \right)\right]
\nonumber \\
&& 
= \frac{ \delta(x- \frac{ {\tilde x}}{2} ) + \delta(x-\frac{ {\tilde x}+1}{2} ) }{2 } \left[ \theta \left({\tilde x} \right) - \theta \left({\tilde x} - 1 \right)\right]
\label{Wbackwarddoubling}
\end{eqnarray}
The last expression means that the backward dynamics is a stochastic Markov chain,
where for each ${\tilde x} \in [0,1[$, one chooses one of the two pre-images 
$\left(x_-=\frac{ {\tilde x}}{2} ; x_+=\frac{ {\tilde x}+1}{2}   \right) $ with the 
equal probabilities $(\frac{1}{2}  , \frac{1}{2})$.

So, even if the probability of a trajectory ${\cal P}_T^{Traj} (x_0,x_1,..,x_T )  $
 in the steady state can be written
either with the forward kernel or with the backward kernel via Eq. \ref{Ptrajforwardback},
one obtains a completely different perspective :

(i) once the initial point $x_0$ is drawn with $ \rho_*(x_0 ) $,
the forward trajectory is then completely determined by the deterministic forward rules $x_{t+1} = f(x_t ) $
\begin{eqnarray}
{\cal P}_T^{Forward(x_0)} (x_1,..,x_T )  =     \prod_{t=0}^{T-1}  w( x_{t+1} \vert x_t \big ) 
= \prod_{t=0}^{T-1}  \delta( x_{t+1} - f(x_t ) \big )
\label{Ptrajforwardx0}
\end{eqnarray}

(ii) once the final point $x_T$ is drawn with $ \rho_*(x_T ) $,
there are $2^T$ possible backward trajectories that all have the same probability $2^{-T}$ 
\begin{eqnarray}
{\cal P}_T^{Backward(x_T)} (x_{T-1},x_{T-2},..,x_0 )  = 
     \prod_{t=0}^{T-1}  W_B( x_t \vert x_{t+1}  \big )
= \prod_{t=0}^{T-1}  \frac{ \delta(x_t- \frac{x_{t+1}}{2} ) + \delta(x_t-\frac{x_{t+1}+1}{2} ) }{2 }
\label{PtrajforwardxT}
\end{eqnarray}


\subsection{ Backward stochastic dynamics for probability density $\rho^B_{\tau}(  x)  $ on the real space 
interval $ x \in [0,1[$}

When considering the backward trajectory, 
it is useful to replace the forward-time $t=0,1,..,T$ by
the backward-time $\tau= T-t  =0,1,..,T$
(see around Eq. \ref{backwardtime} for more details), 
so that the evolution 
for the probability density $\rho^B_{\tau}(  x)  $
in the growing backward time $\tau-1 \to \tau$ 
is governed by the backward kernel of Eq. \ref{Wbackwarddoubling}
\begin{eqnarray}
  \rho^B_{\tau}(  x) = \int_0^1  d {\tilde x}W_B(x \vert  {\tilde x}) \rho^B_{\tau-1}({\tilde x})
= \rho^B_{\tau-1}(2x) \left[ \theta \left( x \right) - \theta \left( x - \frac{1}{2} \right)\right]
+\rho^B_{\tau-1}(2x-1) \left[ \theta \left( x - \frac{1}{2}\right) - \theta \left( x - 1 \right)\right] 
\label{doublingFPbackward}
\end{eqnarray}
The physical interpretation is that the new histogram $\rho^B_{\tau}(  .) $
corresponds to the gluing of two rescaled histograms $\rho^B_{\tau-1}(.) $ on the two intervals $[0,\frac{1}{2}]$ and  
$[\frac{1}{2},1]$.
The iteration of Eq. \ref{doublingFPbackward}
up to the initial density $\rho_{\tau=0}(.)$ at $\tau=0$
\begin{eqnarray}
  \rho^B_{\tau}(  x) 
&&
= \rho^B_{\tau-2}(4x) \left[ \theta \left( x \right) - \theta \left( x - \frac{1}{4} \right)\right]
+\rho^B_{\tau-2}(4x-1) \left[ \theta \left( x - \frac{1}{4}\right) - \theta \left( x -  \frac{1}{2} \right)\right] 
\nonumber \\
&& + \rho^B_{\tau-2}(4x-2) \left[ \theta \left( x -  \frac{1}{2}\right) - \theta \left( x - \frac{3}{4} \right)\right]
+\rho^B_{\tau-2}(4x-3) \left[ \theta \left( x - \frac{3}{4}\right) - \theta \left( x - 1 \right)\right] 
= ...
\nonumber \\
&& =\sum_{j=0}^{2^{\tau}-1} \rho^B_0 \left( 2^{\tau} x - j  \right)
\left[ \theta \left( x -  \frac{j}{2^{\tau}}\right) - \theta \left( x - \frac{j+1}{2^{\tau}} \right)\right]
\label{doublingFPbackwarditer}
\end{eqnarray}
means that the histogram $\rho^B_{\tau}(  .) $ at time $\tau$
corresponds to the gluing of $2^{\tau}$ rescaled histograms  $\rho_{\tau=0}(.) $ on the $2^{\tau}$ intervals $[\frac{j}{2^{\tau}},\frac{j+1}{2^{\tau}}]$ labelled by $j=0,1,..,2^{\tau}-1$.
In particular, the weight of each of these $2^{\tau}$ intervals $[\frac{j}{2^{\tau}},\frac{j+1}{2^{\tau}}]$ is $\frac{1}{2^{\tau}}$
\begin{eqnarray}
\int_{\frac{j}{2^{\tau}}}^{\frac{j+1}{2^{\tau}}} dx   \rho^B_{\tau}(  x) 
&& =\int_{\frac{j}{2^{\tau}}}^{\frac{j+1}{2^{\tau}}} dx 
\rho^B_0 \left( 2^{\tau} x - j  \right)
= \int_0^1 \frac{du}{2^{\tau}} \rho^B_0(u) = \frac{1}{2^{\tau}}
\label{doublingFPbackwarditerintervalle}
\end{eqnarray}
As a consequence, the backward dynamics
will convergence towards the backward uniform steady state
\begin{eqnarray}
\rho^B_{\tau}( {\tilde x}) \opsimeq_{\tau \to + \infty} \rho^B_*( {\tilde x}) =1=\rho_*({\tilde x})
\label{perronFrobsteadyback}
\end{eqnarray}
that coincides with the forward uniform steady state $\rho_*({\tilde x}) =1$ as expected from Eq. \ref{Wbackwardsteady},
but this convergence will be rather weird, since the gluing of $2^{\tau}$ rescaled initial histograms
in Eq. \ref{doublingFPbackwarditer} will introduce discontinuities of the density or of its derivatives 
at all positions $\frac{j}{2^{\tau}} $.


\subsection{ Spectral decomposition of the backward dynamics}

As explained around Eqs \ref{Wbackwardspectral}, 
the spectral decomposition of the backward kernel 
\begin{eqnarray}
w_B(x \vert  {\tilde x})   = \sum_{p=0}^{+\infty} \lambda_p r^B_p(  x)l^B_p( {\tilde x})
\label{Wbackwardspectraldoubling}
\end{eqnarray}
involves the same eigenvalues $\lambda_p=2^{-p}$ of Eq. \ref{eigenvaluesdoubling}
as the forward kernel $w({\tilde x} \vert x)  $,
while the right and left eigenvectors for the backward kernel 
 given by Eq. \ref{eigenvecWB} read in terms of the Bernouilli polynomials of Eq. \ref{bernouilli} 
 and in terms of the singular distributions of Eq. \ref{lndeltasingular}
\begin{eqnarray}
r^B_p(  x) && =l_p(  x)  \rho_*(x) = \frac{ (-1)^{p-1} }{ p!} \left[ \delta^{(p-1)}(x-1) - \delta^{(p-1)}(x)\right] \equiv D_p(x)
\nonumber \\
l^B_p( {\tilde x}) &&  = \frac{ r_p( {\tilde x}) }{ \rho_*({\tilde x})} = B_p({\tilde x})
\label{eigenvecWBdoubling}
\end{eqnarray}
So the right and left eigenvectors have been exchanged with respect to the forward kernel.
The spectral decomposition of Eq. \ref{markovspectral} 
for the backward dynamics
\begin{eqnarray}
\rho^B_{\tau}(x)    =1 +  \sum_{p=1}^{+\infty} 2^{-pt}  D_p(  x) \int_0^1  d {\tilde x}B_p({\tilde x}) \rho^B_0({\tilde x})
\label{perronFrobspectraldoublingbackward}
\end{eqnarray}
is thus very singular, since the distributions $D_p(  x) $ of Eq. \ref{lndeltasingular}
appear directly and not as scalar products as in the forward spectral decomposition of Eq. \ref{perronFrobspectraldoubling}.
These singularities in real space suggest that it is useful ta analyze the backward dynamics
from the point of view of the binary variables as described in the next subsection.


\subsection{ Backward dynamics for the binary variables}

The real-space backward kernel of Eq. \ref{Wbackwarddoubling}
translates for the binary variables $\sigma_l$ of Eq. \ref{binary} into
\begin{eqnarray}
W_B(\sigma_. \vert  {\tilde \sigma }_. )  = \frac{\left( \delta_{ \sigma _1,0} + \delta_{ \sigma _1,1}\right)}{2} \prod_{l=2}^{+\infty} \delta_{\sigma_l,{\tilde \sigma }_{l-1}}
\label{Wdoublingsigmabackward}
\end{eqnarray}
The normalization over $\sigma$ is obvious, 
while the uniform steady density of Eq. \ref{Wdoublingsigmaright} of the forward dynamics
is also the steady density of the backward dynamics
\begin{eqnarray}
P_*^B(  \sigma _. ) = P_*(  \sigma _. )  = \prod_{l=1}^{+\infty} \left( \frac{\delta_{ \sigma _l,0} +  \delta_{\sigma _l,1} }{2} \right)  
\label{Wdoublingsigmarightb}
\end{eqnarray}
by construction (Eq. \ref{Wbackwardsteady}).

The dynamics for the probability $P_{\tau}^B(\sigma_.)$ governed by the backward kernel of Eq. \ref{Wdoublingsigmabackward}
\begin{eqnarray}
P_{\tau}^B(\sigma_1,\sigma_2,\sigma_3,...) && = \sum_{{\tilde \sigma }_.} W_B(\sigma_. \vert  {\tilde \sigma }_. )  P_{\tau-1}^B({\tilde \sigma }_.)
= \frac{\left( \delta_{ \sigma _1,0} + \delta_{ \sigma _1,1}\right)}{2} \prod_{l=2}^{+\infty} \delta_{\sigma_l,{\tilde \sigma }_{l-1}} P_{\tau-1}^B({\tilde \sigma }_.)
\nonumber \\
&& = \frac{\left( \delta_{ \sigma _1,0} + \delta_{ \sigma _1,1}\right)}{2} 
 P_{\tau-1}^B( \sigma _2, \sigma _3,...)
\label{Wdoublingsigmabackwarditer}
\end{eqnarray}
means that one adds the first coefficient $\sigma_1=0,1$ drawn with the probabilities $\left(\frac{1}{2}, \frac{1}{2}\right)$,
while all the other binary coefficients $( \sigma _2, \sigma _3,... )$ are distributed with the shifted probability $P_{\tau-1}^B(.) $
at the previous time $(\tau-1)$.
The iteration up to $\tau=0$ 
\begin{eqnarray}
P_{\tau}^B(\sigma_1,\sigma_2,\sigma_3,...)  
 = \left[ \prod_{l=1}^{\tau} \frac{ \delta_{ \sigma _l,0} + \delta_{ \sigma _l,1}}{2} \right]
 P_0^B( \sigma _{\tau+1}, \sigma _{\tau+2},...)
\label{Wdoublingsigmabackwarditerzero}
\end{eqnarray}
means that the first $\tau$ binary coefficients $\sigma_1=0,1$ are independently 
drawn with the probabilities $\left(\frac{1}{2}, \frac{1}{2}\right)$, while the higher coefficients $l>\tau $
appear with their initial probability $P_0^B( \sigma _{\tau+1}, \sigma _{\tau+2},...) $ at $\tau=0$.
For large $\tau$, Eq. \ref{Wdoublingsigmabackwarditerzero} converges towards the uniform distribution of Eq. \ref{Wdoublingsigmarightb}, but the information on the initial distribution $P_0^B(.) $
is still intact and simply shifted towards the higher coefficients $l>\tau $.


\section{ Trajectory observables via the deformed stochastic backward dynamics }

\label{sec_backwardk}

In this section, the large deviations of trajectory observables
are analyzed via deformations of the stochastic backward dynamics described in the previous section
\ref{sec_backward}.
We stress the similarities and the differences with 
the deformations of the forward deterministic dynamics of section \ref{sec_forwardk}.

\subsection{ Backward dynamics for the generating function via the $k$-deformed backward kernel $ w_B^{[k]}(x \vert  {\tilde x})$  }

As discussed in Appendix \ref{app_backward}, 
the large deviations of trajectories observables
can be alternatively studied via the 
 $k$-deformed backward kernel $w_B(x \vert  {\tilde x}) $ of Eq. \ref{Wbackwardkdeformed}
 that reads for the forward doubling kernel of Eq. \ref{Wdoubling}
 and its uniform steady density $\rho_*(x)=1$
\begin{eqnarray}
 w_B^{[k]}(x \vert  {\tilde x}) &&=   \delta({\tilde x} -f(x) )  e^{ k \Omega (f(x),x)} = \delta({\tilde x} -f(x) )  e^{ k \omega (x)}
\nonumber \\ 
  && =\delta \big({\tilde x} - 2x \big) e^{ k \omega (x)}\left[ \theta \left( x \right) - \theta \left( x - \frac{1}{2} \right)\right]
+\delta \big({\tilde x} - (2x-1) \big) e^{ k \omega (x)}\left[ \theta \left( x - \frac{1}{2}\right) - \theta \left( x - 1 \right)\right]
\nonumber \\
&& 
= \frac{ \delta(x- \frac{ {\tilde x}}{2} ) e^{ k \omega \left(\frac{ {\tilde x}}{2}\right)}+ \delta(x-\frac{ {\tilde x}+1}{2} ) e^{ k \omega \left(\frac{ {\tilde x}+1}{2}\right)}}{2 } 
\equiv  \delta \left(x- \frac{ {\tilde x}}{2} \right) w_{B-}^{[k]}( {\tilde x}) 
+  \delta \left(x-\frac{ {\tilde x}+1}{2} \right) w_{B+}^{[k]}( {\tilde x})
\label{Wbackwardkdeformeddoubling}
\end{eqnarray}
The last line shows that with respect to the undeformed backward kernel $ w_B(x \vert  {\tilde x})$
of Eq. \ref{Wbackward}, 
the two pre-images 
$\left(x_-=\frac{ {\tilde x}}{2} ; x_+=\frac{ {\tilde x}+1}{2}   \right) $ of ${\tilde x}$ 
that were chosen with the equal probabilities $(\frac{1}{2}  , \frac{1}{2})$
are now chosen with the $k$-deformed weights
\begin{eqnarray}
w_{B-}^{[k]}( {\tilde x}) && = \frac{e^{ k \omega \left(\frac{ {\tilde x}}{2}\right)}}{2}
\nonumber \\
w_{B+}^{[k]}( {\tilde x}) && = \frac{e^{ k \omega \left(\frac{ {\tilde x}+1}{2}\right)}}{2}
\label{kweights}
\end{eqnarray}

The backward dynamics for the generating function
\begin{eqnarray}
Z^{B[k]}_{\tau}(x)  && = \int_0^1  d {\tilde x}w^{[k]}_B(x \vert  {\tilde x}) Z^{B[k]}_{\tau-1}({\tilde x})
 \nonumber \\&&
=  
e^{ k \omega (x)} Z^{B[k]}_{\tau-1}(2x) \left[ \theta \left( x \right) - \theta \left( x - \frac{1}{2} \right)\right]
+e^{ k \omega (x)} Z^{B[k]}_{\tau-1}(2x-1) \left[ \theta \left( x - \frac{1}{2}\right) - \theta \left( x - 1 \right)\right]
\label{Wkdeformedmapdoublingiterback}
\end{eqnarray}
corresponds to the $k$-deformation of the backward dynamics of Eq. \ref{doublingFPbackward}
for the probability density $\rho^B_{\tau}(  x)  $.


\subsection{ Eigenvalue problem governing the generating function for large time $\tau \to + \infty$  }

As explained around Eq. \ref{eigenvecWBk}, the 
highest deformed eigenvalue $ \lambda_0^{[k]} $ is the same for the backward and the forward
kernels, while the
the positive right and left eigenvectors of the backward kernel
satisfying
\begin{eqnarray}
\lambda_0^{[k]} \ r_0^{B[k]}(  x) && =  \int d{\tilde x}  \ w^{[k]}_B(x \vert  {\tilde x})  r_0^{B[k]}( {\tilde x})  
=r_0^{B[k]} (2x) e^{ k \omega (x)}\left[ \theta \left( x \right) - \theta \left( x - \frac{1}{2} \right)\right]
+r_0^{B[k]} (2x-1)  e^{ k \omega (x)}\left[ \theta \left( x - \frac{1}{2}\right) - \theta \left( x - 1 \right)\right]
\nonumber \\
\lambda_0^{[k]}  l_0^{B[k]}( {\tilde x}) &&  =  \int d  x  l_0^{B[k]}(  x) w_B^{[k]}(x \vert  {\tilde x})  
=  l_0^{B[k]}\left( \frac{ {\tilde x}}{2} \right)  w_{B-}^{[k]}( {\tilde x}) 
+  l_0^{B[k]}\left( \frac{ {\tilde x}+1}{2} \right)  w_{B+}^{[k]}( {\tilde x})
\nonumber \\
&& =  l_0^{B[k]}\left( \frac{ {\tilde x}}{2} \right)  \frac{e^{ k \omega \left(\frac{ {\tilde x}}{2}\right)}}{2} 
+  l_0^{B[k]}\left( \frac{ {\tilde x}+1}{2} \right)  \frac{e^{ k \omega \left(\frac{ {\tilde x}+1}{2}\right)}}{2}
\label{eigenWkbackdoubling}
\end{eqnarray}
are related to the forward eigenvectors via Eq. \ref{eigenvecWBk}
that reads
 for the doubling map with uniform density $\rho_*(x)=1$
\begin{eqnarray}
r_0^{B[k]}( {\tilde x}) && =l_0^{[k]}( {\tilde x}) 
\nonumber \\
l_0^{B[k]}(  x) &&  =  r_0^{[k]}(  x) 
\label{eigenvecWBkdoubling}
\end{eqnarray}

This exchange between the deformed right and left eigenvectors have very important consequences :

(i) the backward right eigenvector $ r_0^{B[k]}( {\tilde x})  =l_0^{[k]}( {\tilde x}) $
corresponding to the forward left eigenvector
is not a smooth function of $x$.
As a consequence, if one construct 
the probability of the end-point analog to Eq. \ref{probaEndk}
\begin{eqnarray}
P^{B[k]End}_{\tau}(x ) 
\equiv \frac{Z^{B[k]}_{\tau}(x) }{ \int dy Z^{B[k]}_{\tau}(y)  }
\opsimeq_{\tau \to +\infty}
\frac{r_0^{B[k]}(x)}{ \int dy r_0^{B[k]}(y)} = \frac{l_0^{[k]}(x)}{ \int dy l_0^{[k]}(y)}
\equiv P^{B[k]End}_*(x)
 \label{probaEndkZB}
\end{eqnarray}
it will converge towards a singular distribution $P^{[k]End}_*(x) $.

(ii) the backward left eigenvector $ l_0^{B[k]}(  x)   =  r_0^{[k]}(  x) $
corresponding to the forward right eigenvector
is a smooth function of $x$ and can be used to construct the 
canonical conditioned backward kernel as described in the next subsection.


\subsection{ Canonical conditioned backward kernel $w^{C[k]}_B(x \vert  {\tilde x}) $ 
with $k$-deformed probabilities for the two pre-images}

The conditioned backward kernel $w^{C[k]}_B(x \vert  {\tilde x}) $  
of Eq. \ref{WDoobkback}
reads using the backward kernel of Eq. \ref{Wbackwardkdeformeddoubling}
\begin{eqnarray}
w^{C[k]}_B(x \vert  {\tilde x}) 
&& \equiv  \frac{1}{\lambda_0^{[k]}} \  l_0^{B [k]}(x) w^{B[k]}(x \vert  {\tilde x})   \frac{1}{ l_0^{B[k]}({\tilde x})  }
 = \frac{l_0^{B[k]}(x) w^{B[k]}(x \vert  {\tilde x})}{\int d  z  l_0^{B[k]}(  z) w^{B[k]}(z \vert  {\tilde x})  }
\nonumber \\
&& 
  = \frac{  \delta \left(x- \frac{ {\tilde x}}{2} \right)  l_0^{B[k]} \left( \frac{ {\tilde x}}{2} \right)w_{B-}^{[k]}( {\tilde x}) 
+  \delta \left(x-\frac{ {\tilde x}+1}{2} \right)  l_0^{B[k]} \left( \frac{ {\tilde x}+1}{2} \right)w_{B+}^{[k]}( {\tilde x})}
{  l_0^{B[k]} \left( \frac{ {\tilde x}}{2} \right) w_{B-}^{[k]}( {\tilde x}) 
+  l_0^{B[k]} \left( \frac{ {\tilde x}+1}{2} \right) w_{B+}^{[k]}( {\tilde x})  }
\nonumber \\
&& \equiv  \delta \left(x- \frac{ {\tilde x}}{2} \right) p_{B-}^{C[k]}( {\tilde x}) 
+  \delta \left(x-\frac{ {\tilde x}+1}{2} \right) p_{B+}^{C[k]}( {\tilde x})
  \label{WDoobkbackdoubling}
\end{eqnarray}
The last line shows that with respect to the undeformed backward kernel $ w_B(x \vert  {\tilde x})$
of Eq. \ref{Wbackward}, 
the two pre-images 
$\left(x_-=\frac{ {\tilde x}}{2} ; x_+=\frac{ {\tilde x}+1}{2}   \right) $ of ${\tilde x}$ 
that were chosen with the equal probabilities $(\frac{1}{2}  , \frac{1}{2})$
are now chosen with the $k$-deformed complementary probabilities $p_{B-}^{C[k]}( {\tilde x})+p_{B+}^{C[k]}( {\tilde x})=1 $
that involve the deformed weights $w_{B\pm}^{[k]}(.) $ of Eq. \ref{kweights}
and the deformed backward left eigenvector $ l_0^{B[k]}(.) $ of Eq. \ref{eigenWkbackdoubling}
\begin{eqnarray}
p_{B-}^{C[k]}( {\tilde x}) && = 
\frac{    l_0^{B[k]} \left( \frac{ {\tilde x}}{2} \right)w_{B-}^{[k]}( {\tilde x}) }
{  l_0^{B[k]} \left( \frac{ {\tilde x}}{2} \right)  w_{B-}^{[k]}( {\tilde x}) 
+  l_0^{B[k]} \left( \frac{ {\tilde x}+1}{2} \right) w_{B+}^{[k]}( {\tilde x})  }
= \frac{    l_0^{B[k]} \left( \frac{ {\tilde x}}{2} \right) e^{ k \omega \left(\frac{ {\tilde x}}{2}\right)} }
{  l_0^{B[k]} \left( \frac{ {\tilde x}}{2} \right)  e^{ k \omega \left(\frac{ {\tilde x}}{2}\right)} 
+  l_0^{B[k]} \left( \frac{ {\tilde x}+1}{2} \right) e^{ k \omega \left(\frac{ {\tilde x}+1}{2}\right)}  }
\nonumber \\
p_{B+}^{C[k]}( {\tilde x}) &&  = 
\frac{    l_0^{B[k]} \left( \frac{ {\tilde x}+1}{2} \right)w_{B+}^{[k]}( {\tilde x})}
{  l_0^{B[k]} \left( \frac{ {\tilde x}}{2} \right)w_{B-}^{[k]}( {\tilde x}) 
+  l_0^{B[k]} \left( \frac{ {\tilde x}+1}{2} \right) w_{B+}^{[k]}( {\tilde x})  } 
 = 
\frac{    l_0^{B[k]} \left( \frac{ {\tilde x}+1}{2} \right) e^{ k \omega \left(\frac{ {\tilde x}+1}{2}\right)}}
{  l_0^{B[k]} \left( \frac{ {\tilde x}}{2} \right) e^{ k \omega \left(\frac{ {\tilde x}}{2}\right)} 
+  l_0^{B[k]} \left( \frac{ {\tilde x}+1}{2} \right) e^{ k \omega \left(\frac{ {\tilde x}+1}{2}\right)}  } 
\label{Ckprobadoubling}
\end{eqnarray}

The delta-function identity yields that Eq. \ref{WDoobkbackdoubling}
can be rewritten as
\begin{eqnarray}
w^{C[k]}_B(x \vert  {\tilde x}) &&  \equiv  \delta \left(x- \frac{ {\tilde x}}{2} \right) p_{B-}^{C[k]}( {\tilde x}) 
+  \delta \left(x-\frac{ {\tilde x}+1}{2} \right) p_{B+}^{C[k]}( {\tilde x})
\nonumber \\
&& = \delta \left({\tilde x}-2x \right) p_{B-}^{C[k]}( {\tilde x}) 
\left[  \theta \left( x \right) -  \theta \left( x - \frac{1}{2} \right) \right]
+  \delta \left({\tilde x}-(2x-1)  \right) p_{B+}^{C[k]}( {\tilde x})
\left[ \theta \left(x - \frac{1}{2} \right) - \theta \left(x - 1 \right)\right]
  \label{WDoobkbackdoublingfinal}
\end{eqnarray}

The conditioned backward steady state $\rho^{BC[k]}_*( x)  $ corresponding its right eigenvector
associated to the eigenvalue unity 
\begin{eqnarray}
\rho^{BC[k]}_*( x) && = \int_0^1 d{\tilde x} w^{C[k]}_B(x \vert  {\tilde x}) \rho^{BC[k]}_*( {\tilde x})
\nonumber \\
&& =  p_{B-}^{C[k]}( 2x) \rho^{BC[k]}_*( 2x)
\left[  \theta \left( x \right) -  \theta \left( x - \frac{1}{2} \right) \right]
+  p_{B+}^{C[k]}( 2x-1) \rho^{BC[k]}_*( 2x-1)
\left[ \theta \left(x - \frac{1}{2} \right) - \theta \left(x - 1 \right)\right]
  \label{rhoDoobkbackdoublingfinal}
\end{eqnarray}
is given by Eq. \ref{rhosteadyDoobbackward}
\begin{eqnarray}
\rho^{BC[k]}_*( x) \equiv l_0^{B[k]}(  x) r_0^{B[k]}(  x)
=  l_0^{[k]}(x) r_0^{[k]}(x) 
  \label{rhosteadyDoobbackwardprod}
  \end{eqnarray}
and is thus expected to be singular in $x$.
It is thus useful to consider the formulation of the backward deformed and conditioned dynamics
in terms of the binary variables.


\subsection{ Analysis via the backward deformed and conditioned dynamics for the binary variables $\sigma_l $ }

\subsubsection{ Properties of the backward deformed kernel in the binary variables $\sigma_l $ }

With the notation
\begin{eqnarray}
\Omega(\sigma_.) \equiv \omega \left( x=\sum_{l=1}^{+\infty} \frac{\sigma_l}{2^l}  \right)
\label{Omegasigma}
\end{eqnarray}
for the parametrization of the trajectory observable of Eq. \ref{observablemap}
in terms of the binary observables $\sigma_l$,
the $k$-deformation of the backward kernel of Eq. \ref{Wdoublingsigmabackward} 
\begin{eqnarray}
W^{[k]}_B(\sigma_. \vert  {\tilde \sigma }_. )  
&& = e^{k \Omega(\sigma_.) } W_B(\sigma_. \vert  {\tilde \sigma }_. ) 
=e^{k \Omega(\sigma_.) } \frac{\left( \delta_{ \sigma _1,0} + \delta_{ \sigma _1,1}\right)}{2} \prod_{l=2}^{+\infty} \delta_{\sigma_l,{\tilde \sigma }_{l-1}}
\nonumber \\
&& \equiv \left[ \delta_{ \sigma _1,0} W_{B-}^{[k]}({\tilde \sigma } )
+ \delta_{ \sigma _1,1} W_{B+}^{[k]}({\tilde \sigma } )\right] \prod_{l=2}^{+\infty} \delta_{\sigma_l,{\tilde \sigma }_{l-1}}
  \label{Wkbackbinary}
\end{eqnarray}
involve the two following weights are the analog of the real-space weights of Eq. \ref{kweights}
\begin{eqnarray}
W_{B-}^{[k]}({\tilde \sigma } )&& = \frac{e^{ k \Omega (0,{\tilde \sigma }_1,{\tilde \sigma }_2,..)}}{2}
\nonumber \\
W_{B+}^{[k]}({\tilde \sigma } ) && = \frac{e^{ k \Omega ( 1,{\tilde \sigma }_1,{\tilde \sigma }_2,..) }}{2}
\label{kweightsigma}
\end{eqnarray}

The eigenvalues equations for the backward deformed eigenvectors
read
\begin{eqnarray}
\lambda_0^{[k]} L^{B[k]}_0({\tilde \sigma }_1,{\tilde \sigma }_2,...)  
&& = \sum_{\sigma_.}L^{B[k]}_0(\sigma_.)W^{[k]}_B(\sigma_. \vert  {\tilde \sigma }_. )   
\nonumber \\
&& = 
 \frac{L^{B[k]}_0(0,{\tilde \sigma }_1,{\tilde \sigma }_2,..)
e^{k \Omega(0,{\tilde \sigma }_1,{\tilde \sigma }_2,..) }
+ L^{B[k]}_0(1,{\tilde \sigma }_1,{\tilde \sigma }_2,..)
e^{k \Omega(1,{\tilde \sigma }_1,{\tilde \sigma }_2,..) }}{2} 
\label{Wdoublingsigmabackwardkleft}
\end{eqnarray}
and
\begin{eqnarray}
\lambda_0^{[k]} R^{B[k]}_0( \sigma _1, \sigma _2,...)  
&& = \sum_{\tilde \sigma }W^{[k]}_B(\sigma_. \vert  {\tilde \sigma }_. )   R^{B[k]}_0({\tilde \sigma }_.)
\nonumber \\
&& = 
 \frac{ \delta_{ \sigma _1,0}
e^{k \Omega(0,\sigma_2,\sigma_3,..) }
+  \delta_{ \sigma _1,1}
e^{k \Omega(1,\sigma_2,\sigma_3,..) }}{2} R^{B[k]}_0(\sigma_2,\sigma_3,...)
\label{Wdoublingsigmabackwardkright}
\end{eqnarray}


\subsubsection{ Construction of the backward conditioned kernel in the binary variables $\sigma_l $ }

The deformed left eigenvector $L^{B[k]}_0(\sigma_.) $ of Eq. \ref{Wdoublingsigmabackwardkleft}
is useful to construct the backward conditioned kernel via Eq. \ref{WDoobkback}
\begin{eqnarray}
W^{C[k]}_B(\sigma_. \vert  {\tilde \sigma }_. )  
&& \equiv  \frac{1}{\lambda_0^{[k]}} \ L^{B[k]}_0(\sigma_.) W^{[k]}_B(\sigma_. \vert  {\tilde \sigma }_. )    \frac{1}{  L^{B[k]}_0({\tilde \sigma }_.)  }
= \frac{ L^{B[k]}_0(\sigma_.) W^{[k]}_B(\sigma_. \vert  {\tilde \sigma }_. )}{
\sum_{\sigma_.'}  L^{B[k]}_0(\sigma_.') W^{[k]}_B(\sigma_.' \vert  {\tilde \sigma }_. )}
\nonumber \\
&& =\left[ \delta_{ \sigma _1,0} p_{B-}^{C[k]}({\tilde \sigma }_. )
+ \delta_{ \sigma _1,1} p_{B+}^{C[k]}({\tilde \sigma }_. )\right] \prod_{l=2}^{+\infty} \delta_{\sigma_l,{\tilde \sigma }_{l-1}}
  \label{WDoobkbackbinary}
\end{eqnarray}
where the two complementary probabilities $ p_{B-}^{C[k]}({\tilde \sigma }_. )+p_{B+}^{C[k]}({\tilde \sigma }_. )=1$ read
\begin{eqnarray}
p_{B-}^{C[k]}({\tilde \sigma }_. ) && \equiv \frac{L^{B[k]}_0(0,{\tilde \sigma }_1,{\tilde \sigma }_2,..)
e^{k \Omega(0,{\tilde \sigma }_1,{\tilde \sigma }_2,..) }}{L^{B[k]}_0(0,{\tilde \sigma }_1,{\tilde \sigma }_2,..)
e^{k \Omega(0,{\tilde \sigma }_1,{\tilde \sigma }_2,..) }
+ L^{B[k]}_0(1,{\tilde \sigma }_1,{\tilde \sigma }_2,..)
e^{k \Omega(1,{\tilde \sigma }_1,{\tilde \sigma }_2,..)}}
\nonumber \\
p_{B+}^{C[k]}({\tilde \sigma }_. ) && \equiv \frac{L^{B[k]}_0(1,{\tilde \sigma }_1,{\tilde \sigma }_2,..)
e^{k \Omega(1,{\tilde \sigma }_1,{\tilde \sigma }_2,..) }}{L^{B[k]}_0(0,{\tilde \sigma }_1,{\tilde \sigma }_2,..)
e^{k \Omega(0,{\tilde \sigma }_1,{\tilde \sigma }_2,..) }
+ L^{B[k]}_0(1,{\tilde \sigma }_1,{\tilde \sigma }_2,..)
e^{k \Omega(1,{\tilde \sigma }_1,{\tilde \sigma }_2,..)}}
\label{probaBCbinary}
\end{eqnarray}

This formulation of the backward conditioned dynamics in the binary variables $\sigma_l$
is indeed helpful to clarify the properties that are singular in the real-space variable $x \in [0,1[$,
as shown explicitly on an example of trajectory observable in the next section.


\section{ Example of the trajectory observable $\omega(x)=x$ from various perspectives }

\label{sec_omegax}

In this section, the different perspectives 
described in sections \ref{sec_forwardk} and \ref{sec_backwardk}
are illustrated on the simplest trajectory observable $\omega(x)=x $ for the doubling map \cite{naftali}.


\subsection{ Analysis via the $k$-deformed forward dynamics on the real-space interval $x \in [0,1[$ }

\label{subsec_omegaxlinear}

\subsubsection{ Exact solution for the deformed eigenvalue $\lambda_0^{[k]} $ and 
the deformed right eigenvector $r_0^{[k]}(.) $ \cite{naftali}} 

The example of the trajectory observable $\omega(x)=x$
where the right eigenvalue Eq. \ref{eigenWkmapdoubling} reads
\begin{eqnarray}
\lambda_0^{[k]}  r_0^{[k]}( {\tilde x}) 
=   \frac{  e^{ k \frac{ {\tilde x}}{2}} r_0^{[k]}(  \frac{ {\tilde x}}{2})
 +  e^{ k \frac{ {\tilde x}+1}{2}} r_0^{[k]}(  \frac{ {\tilde x}+1}{2})}{2 }
\label{eigenrightdoublingx}
\end{eqnarray}
was solved in \cite{naftali} with the $k$-deformed eigenvalue
\begin{eqnarray}
\lambda_0^{[k]} = \frac{1+e^k}{2} 
\label{lambdadoublingx}
\end{eqnarray}
and the $k$-deformed right eigenvector
\begin{eqnarray}
  r_0^{[k]}( {\tilde x}) =  e^{k \left({\tilde x} - \frac{1}{2} \right)} 
\label{rightdoublingx}
\end{eqnarray}

The corresponding left eigenvalue Eq. \ref{eigenWkmapdoubling} 
\begin{eqnarray}
0 \leq x < \frac{1}{2} : \ \ 
\lambda_0^{[k]}  l_0^{[k]}(  x) &&  = l_0^{[k]}( 2x ) e^{ k x } 
\nonumber \\
\frac{1}{2} \leq x < 1 : \ \ 
\lambda_0^{[k]}  l_0^{[k]}(  x) &&  = l_0^{[k]}(2x-1) e^{ k x }
\label{eigenWkmapdoublingleft}
\end{eqnarray}
is expected to be singular in real space as soon as $k \ne 0$,
as indicated by the perturbation theory in the next subsection.


\subsubsection{ Comparison with the general perturbation theory in $k$ }

Let us compare with the perturbation theory of subsection \ref{subsec_perturbation}.
The series expansion of the exact eigenvalue of Eq. \ref{lambdadoublingx}
\begin{eqnarray}
\lambda_0^{[k]} = \frac{1+e^k}{2} =  1+ \frac{k}{2} + \frac{ k^2 }{4} +O(k^3)
\label{lambdadoublingxper}
\end{eqnarray}
is in agreement with the first-order correction of Eq \ref{lambdaper1doubling}
and the second-order correction Eq. \ref{lambdaper2doubling}
\begin{eqnarray}
  \lambda^{(1)} && =  \int_0^1dx  x = \frac{1}{2}
\nonumber \\
 \lambda^{(2)}  && =   \int_0^1dx  \frac{ x^2} {2}     +   \int_0^1dx  x  \left( x - \frac{1}{2} \right)
 = \frac{1}{4}
\label{lambdaper2check}
\end{eqnarray}
The series expansion of the exact right eigenvector of Eq. \ref{rightdoublingx}
\begin{eqnarray}
  r_0^{[k]}( {\tilde x}) =  e^{k \left({\tilde x} - \frac{1}{2} \right)} =1+ k \left({\tilde x} - \frac{1}{2} \right) + O(k^2)
\label{rightdoublingxper}
\end{eqnarray}
is in agreement with the first-order correction $r^{(1)}_0 ({\tilde x})  $ 
Eq. \ref{eigenvecper1rightdoubling} for $\omega(x)=x$
that reduces to the first Bernoulli polynomial $B_1({\tilde x})$ of Eq. \ref{bernouilli} 
\begin{eqnarray}
   r^{(1)}_0 ({\tilde x}) =   B_1 ({\tilde x}) =\left({\tilde x} - \frac{1}{2} \right)
   \label{eigenvecper1rightdoublingpercheck}
\end{eqnarray}

The first-order perturbation theory of Eq. \ref{eigenvecper1leftdoubling} for the left eigenvector
\begin{eqnarray}
  l^{(1)}_0 ({\tilde x})    = \sum_{p=1}^{+\infty}  D_p ({\tilde x}) 
  \left[ \frac{ 1 }{1- 2^{-p}}  \int_0^1dx  x  B_p (x) \right]
  =  \sum_{p=1}^{+\infty}  D_p ({\tilde x}) 
  \left[ \frac{ 1 }{1- 2^{-p}}  \int_0^1dx  B_1(x)  B_p (x) \right]
\label{eigenvecper1leftdoublingx}
\end{eqnarray}
 involves the singular distributions $D_p(.)$ of Eq. \ref{lndeltasingular} for all $p=1,2,..,+\infty$.


\subsection{ Analysis via the $k$-deformed forward kernel for the binary variables $\sigma_l $ }

For $\omega(x)=x$, the $k$-deformed kernel 
\begin{eqnarray}
W^{[k]}( {\tilde \sigma }_. \vert \sigma_.)  \equiv e^{k \Omega(\sigma_.) }\prod_{l=1}^{+\infty} \delta_{{\tilde \sigma }_l,\sigma_{l+1}}
\label{Wdoublingsigmak}
\end{eqnarray}
is factorized in the binary variables
\begin{eqnarray}
W^{[k]}( {\tilde \sigma }_. \vert \sigma_.)  =
 e^{ \displaystyle k  \sum_{l=1}^{+\infty} \frac{\sigma_l}{2^l}  }\prod_{l=1}^{+\infty} \delta_{{\tilde \sigma }_l,\sigma_{l+1}}
 = e^{k  \frac{\sigma_1}{2}  }\prod_{l=1}^{+\infty} \left[e^{k  \frac{\sigma_{l+1}}{2^{l+1}}  }\ \delta_{{\tilde \sigma }_l,\sigma_{l+1}} \right]
\label{Wdoublingsigmakx}
\end{eqnarray}

The $k$-deformed right eigenvector of Eq. \ref{rightdoublingx}
in real space translates for the binary variables into the factorized form
\begin{eqnarray}
R_0^{[k]}(\sigma_.) 
  = e^{- \frac{k}{2} } \prod_{l=1}^{+\infty} \frac{ e^{k  \frac{\sigma_{l}}{2^{l}}  } }{2} 
\label{r0sigmakx}
\end{eqnarray}

The solution for the corresponding left eigenvector 
\begin{eqnarray}
 \frac{1+e^k}{2} \  L_0^{[k]}(\sigma_1,\sigma_2,...) 
 && = \sum_{{\tilde \sigma }_.}  L_0^{[k]}({\tilde \sigma }) W^{[k]}( {\tilde \sigma }_. \vert \sigma_.)  
 = \sum_{{\tilde \sigma }_.}  L_0^{[k]}({\tilde \sigma })
 e^{k  \frac{\sigma_1}{2}  }\prod_{l=1}^{+\infty} \left[e^{k  \frac{\sigma_{l+1}}{2^{l+1}}  }\ \delta_{{\tilde \sigma }_l,\sigma_{l+1}} \right]
 \nonumber \\
&& =  L_0^{[k]}(\sigma_2,\sigma_3,...) \prod_{l=1}^{+\infty}e^{k  \frac{\sigma_l}{2^l}  }
\label{Wdoublingsigmakxleft}
\end{eqnarray}
will be easier to derive later with its appropriate normalization via the backward perspective
(see Eq. \ref{l0sigmakx}).


\subsection{ Analysis via the $k$-conditioned backward kernel $w^{C[k]}_B(x \vert  {\tilde x}) $ on the real-space interval $x \in [0,1[$ }

The eigenvalue $\lambda_0^{[k]} = \frac{1+e^k}{2}  $
of Eq. \ref{lambdadoublingx}
that is associated to the $k$-deformed forward right eigenvector of Eq. \ref{rightdoublingx}
is now associated to the $k$-deformed backward left eigenvector of Eq. \ref{eigenvecWBkdoubling}
\begin{eqnarray}
  l_0^{B[k]}( {\tilde x})  = r_0^{[k]}( {\tilde x}) =  e^{k \left({\tilde x} - \frac{1}{2} \right)} 
\label{leftBdoublingx}
\end{eqnarray}
As a consequence, the two $k$-deformed probabilities of Eq. \ref{Ckprobadoubling} 
\begin{eqnarray}
p_{B-}^{C[k]}( {\tilde x}) 
&&=\frac{1}{1+e^k} 
\nonumber \\
p_{B+}^{C[k]}( {\tilde x}) 
&&   = \frac{e^k}{1+e^k} 
\label{Ckprobadoublingx}
\end{eqnarray}
are independent of ${\tilde x}$,
and the conditioned backward kernel of Eq. \ref{WDoobkbackdoublingfinal} reduces to
\begin{eqnarray}
w^{C[k]}_B(x \vert  {\tilde x}) && 
= \frac{  \delta \left(x- \frac{ {\tilde x}}{2} \right) +  \delta \left(x-\frac{ {\tilde x}+1}{2} \right) e^k}
{1+e^k}
\nonumber \\
&& = \frac{ \delta \left({\tilde x}-2x \right) }{1+e^k}  \left[ \theta \left( x \right) - \theta \left( x - \frac{1}{2} \right)\right]
+  \frac{ \delta \left({\tilde x}-(2x-1)  \right) e^k} {1+e^k} \left[ \theta \left( x - \frac{1}{2}\right) - \theta \left( x - 1 \right)\right]
  \label{WDoobkbackdoublingx}
\end{eqnarray}

However the conditioned backward steady state $\rho^{BC[k]}_*( x) $ of Eq. \ref{rhosteadyDoobbackwardprod}
 is expected to be singular in $x$, and it is thus useful to turn to 
the formulation of the backward conditioned dynamics
in terms of the binary variables.


\subsection{ Analysis via the backward deformed and conditioned kernel for the binary variables $\sigma_l $ }

\subsubsection{ Backward deformed kernel for the binary variables $\sigma_l $ }

For $\omega(x)=x$, the backward deformed kernel of Eq. \ref{Wkbackbinary}
is factorized in the binary variables
\begin{eqnarray}
W^{[k]}_B(\sigma_. \vert  {\tilde \sigma }_. )  
=e^{ \displaystyle k  \sum_{l=1}^{+\infty} \frac{\sigma_l}{2^l}  }
\left( \frac{ \delta_{ \sigma _1,0} + \delta_{ \sigma _1,1}}{2}\right) \prod_{l=2}^{+\infty} \delta_{\sigma_l,{\tilde \sigma }_{l-1}}
 = 
\left( \frac{ \delta_{ \sigma _1,0} + e^{  \frac{k}{2  }} \delta_{ \sigma _1,1}}{2} \right)
 \prod_{l=2}^{+\infty} \left[ e^{  k  \frac{\sigma_l}{2^l}  } \delta_{\sigma_l,{\tilde \sigma }_{l-1}} \right]
\label{Wdoublingsigmabackwardkx}
\end{eqnarray}

The backward left eigenvector of Eq. \ref{leftBdoublingx} in real space
translates for the binary variables into
\begin{eqnarray}
L_0^{B[k]}(\sigma_.) 
  = e^{- \frac{k}{2}} \prod_{l=1}^{+\infty}  e^{k  \frac{\sigma_l}{2^l } }
\label{lB0sigmakx}
\end{eqnarray}


\subsubsection{ Backward conditioned kernel for the binary variables $\sigma_l $ with its explicit conditioned
steady density $P^{C[k]}_B(\sigma_.  ) $}

The backward conditioned kernel of Eq. \ref{WDoobkbackdoublingx}
in real space translates for the binary variables into
\begin{eqnarray}
W^{C[k]}_B(\sigma_. \vert  {\tilde \sigma }_. )  
= \left( \frac{ \delta_{ \sigma _1,0} + e^k \delta_{ \sigma _1,1}}{ 1+e^k} \right) \prod_{l=2}^{+\infty} \delta_{\sigma_l,{\tilde \sigma }_{l-1}}
\label{WdoublingsigmabackwardkxConditioned}
\end{eqnarray}
The corresponding conditioned steady probability is now obvious 
and factorized in these binary variables
\begin{eqnarray}
P^{C[k]}_B(\sigma_.  )  
=\prod_{l=1}^{+\infty} \left( \frac{ \delta_{ \sigma _l,0} + e^k \delta_{ \sigma _l,1}}{ 1+e^k} \right)
= \prod_{l=1}^{+\infty} \left( \frac{  e^{ k \sigma_l} }{ 1+e^k} \right)
\label{PdoublingsigmabackwardkxConditioned}
\end{eqnarray}

Since the translation in real-space $x \in [0,1[$ 
\begin{eqnarray}
\rho^{C[k]}_B(x)   
= \sum_{\sigma_.}  P^{C[k]}_B(  \sigma_.)  \delta \left( x-\displaystyle \sum_{l=1}^{+\infty} \frac{\sigma_l}{2^l}\right)
\label{PdoublingsigmabackwardkxConditionedx}
\end{eqnarray}
is not smooth, it is useful to characterize this real-space density $\rho^{C[k]}_B(x) $
via its Fourier coefficients of Eq. \ref{rhofouriercoefs} for $n \in \mathbf{Z}$ 
\begin{eqnarray}
 {\hat \rho}^{C[k]}_B   (n ) &&= \int_0^1 dx \rho^{C[k]}_B(x)    e^{- i 2 \pi n x} 
 = \sum_{\sigma_.}  P^{C[k]}_B(  \sigma_.)  e^{ \displaystyle - i 2 \pi n \sum_{l=1}^{+\infty} \frac{\sigma_l}{2^l} } 
= 
 \prod_{l=1}^{+\infty} \left( \sum_{\sigma_l=0,1} \frac{  e^{ \left( k - i 2 \pi \frac{n}{2^l}\right) \sigma_l} }{ 1+e^k} \right)
 \nonumber \\
&& 
 =  \prod_{l=1}^{+\infty} \left(  \frac{ 1+ e^{ \left( k - i 2 \pi \frac{n}{2^l}\right) } }{ 1+e^k} \right)
\label{rhoBCKfouriercoefs}
\end{eqnarray}


\subsubsection{ Explicit expression of the backward deformed right eigenvector $R_0^{B[k]}(\sigma_.) $  }

Since the backward conditioned steady state $P^{C[k]}_B(\sigma_.  )  $ 
given by the explicit expression of Eq. \ref{PdoublingsigmabackwardkxConditioned}
coincides with the product 
 $L_0^{B[k]}(\sigma_.)  R_0^{B[k]}(\sigma_.)  $ of the backward
 deformed eigenvectors, we can use the explicit expression of Eq. \ref{lB0sigmakx}
  for the left eigenvector $L_0^{B[k]}(\sigma_.) $ 
  to obtain that the right eigenvector $R_0^{B[k]}(\sigma_.)  $ reads
\begin{eqnarray}
R_0^{B[k]}(\sigma_.) = \frac{P^{C[k]}_B(\sigma_.  )  }{L_0^{B[k]}(\sigma_.) } 
  = \prod_{l=1}^{+\infty} \frac{ \left( \frac{ \delta_{ \sigma _l,0} + e^k \delta_{ \sigma _l,1}}{ 1+e^k} \right)}
  {   e^{k  \frac{\sigma_l}{2^l}}   }
  = \prod_{l=1}^{+\infty} \left[ \frac{\left( \delta_{ \sigma _l,0} + e^{  k (1-\frac{1}{2^l} ) } \delta_{\sigma_l,1} \right)}{1+e^k}  \right]
\label{rB0sigmakx}
\end{eqnarray}
and satisfies indeed the eigenvalue equation involving the backward deformed kernel 
 of Eq. \ref{Wdoublingsigmabackwardkx}
\begin{eqnarray}
\sum_{{\tilde \sigma}_.}  W^{[k]}_B(\sigma_. \vert  {\tilde \sigma }_. )  R_0^{B[k]}({\tilde \sigma}_.)
= \frac{1+e^k}{2} R_0^{B[k]}( \sigma_.)
\label{Wdoublingsigmabackwardkxright}
\end{eqnarray}

Since the translation in real-space $x \in [0,1[$ 
\begin{eqnarray}
r_0^{B[k]}(x) 
= \sum_{\sigma_.}  R_0^{B[k]}(\sigma_.)  \delta \left( x-\displaystyle \sum_{l=1}^{+\infty} \frac{\sigma_l}{2^l}\right)
\label{r0bx}
\end{eqnarray}
is not smooth, it is useful to to characterize this real-space eigenvector $r_0^{B[k]}(x)  $
via its Fourier coefficients of Eq. \ref{rhofouriercoefs} for $n \in \mathbf{Z}$ 
\begin{eqnarray}
 {\hat r}_0^{B[k]}   (n ) &&= \int_0^1 dx r_0^{B[k]}(x)     e^{- i 2 \pi n x} 
 = \sum_{\sigma_.}  r_0^{B[k]}(\sigma_.)  e^{ \displaystyle - i 2 \pi n \sum_{l=1}^{+\infty} \frac{\sigma_l}{2^l} } 
\nonumber \\
&& 
 =  \prod_{l=1}^{+\infty} \left(  \frac{ 1+ e^{ \left[ k(1-\frac{1}{2^l} ) - i 2 \pi \frac{n}{2^l}\right] } }{ 1+e^k} \right)
\label{r0bxfouriercoefs}
\end{eqnarray}


\subsubsection{ Explicit expression of the forward deformed left eigenvector $L_0^{[k]}(\sigma_.) $  }

Similarly, the forward deformed left eigenvector $L_0^{[k]}(\sigma_.) $
can be computed from the ratio between the conditioned density $ \rho^{C[k]}(\sigma_.  )=\rho^{C[k]}_B(\sigma_.  )$ of Eq. \ref{PdoublingsigmabackwardkxConditioned}
and the forward deformed right eigenvector $R_0^{[k]}(\sigma_.) $ of Eq. \ref{r0sigmakx},
\begin{eqnarray}
L_0^{[k]}(\sigma_.) = \frac{P^{C[k]}(\sigma_.  )}{R_0^{[k]}(\sigma_.)}
  = \prod_{l=1}^{+\infty} \frac{ \left( \frac{ \delta_{ \sigma _l,0} + e^k \delta_{ \sigma _l,1}}{ 1+e^k} \right)}
  { \left(  \frac{ e^{k  \frac{\sigma_{l}}{2^{l}}  } }{2} \right)}
  = \prod_{l=1}^{+\infty} \left[ \frac{2}{1+e^k} \left( \delta_{ \sigma _l,0} + e^{  k (1-\frac{1}{2^l} ) } \delta_{\sigma_l,1} \right) \right]
  = \prod_{l=1}^{+\infty} \left[ \frac{2 e^{  k \sigma_l (1-\frac{1}{2^l} ) }}{1+e^k}  \right]
\label{l0sigmakx}
\end{eqnarray}
that satisfies indeed the eigenvalue Eq. \ref{Wdoublingsigmakxleft}.

In real space, the forward deformed left eigenvector $ l_0^{[k]}(x) = r_0^{B[k]}(x)  $
coincides with the backward deformed right eigenvector $r_0^{B[k]}(x) $ of Eq. \ref{r0bx}
with its Fourier coefficients of Eq. \ref{r0bxfouriercoefs} :  
the singular character of $ l_0^{[k]}(x) $ in real space
was already found numerically on Fig S1 of \cite{spain}.


\section{ Large deviations for the backward empirical density in the binary space }

\label{sec_level2}

In this section, the explicit large deviations at level 2.5 for Markov chains
(see the reminder in subsection \ref{subsec_higherlevels} of Appendix A)
are applied to the stochastic backward dynamics associated to the doubling map
in order to obtain the explicit large deviations at level 2 for the empirical density alone.

\subsection{ Reminder on the explicit large deviations at level 2.5 for an arbitrary Markov chain
in binary space }

For an arbitrary Markov chain in binary space governed by the kernel $W(\sigma_. \vert  {\tilde \sigma}_.) $,
the joint distribution $P^{[2.5]}_T \left[ {\mathring P} (.) ; {\mathring W}(.\vert.) \right] $ of the empirical probability ${\mathring P}(\sigma_.)$
and of the empirical kernel ${\mathring W}(\sigma_. \vert  {\tilde \sigma}_.) $ 
seen during a trajectory over the time-window $T$ follows the large deviation form
at the so-called level 2.5 (see the reminder in Appendix A around Eq. \ref{pnormaempi} in real-space variables)
\begin{eqnarray}
P^{[2.5]}_T \left[ {\mathring P} (.) ; {\mathring W}(.\vert.) \right] 
&& \opsimeq_{T \to + \infty}
   \delta \left( \sum_{\sigma_.} {\mathring P}(\sigma_.)  - 1 \right) 
  \left[ \prod_{ {\tilde \sigma}_.} \delta \left(  \sum_{\sigma_.}  {\mathring W}(\sigma_. \vert  {\tilde \sigma}_.) - 1 \right) \right]
 \left[\prod_{ \sigma_.} \delta \left(  \sum_{{\tilde \sigma}_.}   {\mathring W}(\sigma_. \vert  {\tilde \sigma}_.) {\mathring P}({\tilde \sigma}_.)- {\mathring P}(\sigma_.) \right)  \right] \ \nonumber \\
 && e^{\displaystyle -T  I_{2.5}\left[ {\mathring P} (.) ; {\mathring W}(.\vert.)\right] }
\label{level2.5chain}
\end{eqnarray}
where the rate function $I_{2.5}\left[ {\mathring P} (.) ; {\mathring W}(.\vert.)\right] $ at level 2.5 
is given by the explicit relative entropy
\begin{eqnarray}
  I_{2.5} \left[ {\mathring P} (.) ; {\mathring W}(.\vert.)\right]
  =   \sum_{\sigma_.}\sum_{{\tilde \sigma}_.} 
    {\mathring W}(\sigma_. \vert  {\tilde \sigma}_.){\mathring P}({\tilde \sigma}_.)
  \ln \left( \frac{{\mathring W}(\sigma_. \vert  {\tilde \sigma}_.)}{ W(\sigma_. \vert  {\tilde \sigma}_.)  }  \right) 
\label{rate2.5chain}
\end{eqnarray}
while the prefactors in front of the exponential in Eq. \ref{level2.5chain}
impose
the three following constitutive constraints for the empirical observables
(see the reminder of Appendix A around Eqs \ref{rho1ptnorma} \ref{empiwnorma} \ref{empiwsteady}
in real-space variables)

(i)  the empirical probability ${\mathring P}(\sigma_.)$ has to be normalized over $\sigma_. $  

(ii) the empirical kernel ${\mathring W}(\sigma_. \vert  {\tilde \sigma}_.) $
has to be normalized over $\sigma_. $ for any ${\tilde \sigma}_. $

(iii) the empirical probability ${\mathring P}(\sigma_.)$ should be steady with respect to the 
dynamics governed by
the empirical kernel ${\mathring W}(\sigma_. \vert  {\tilde \sigma}_.) $.

As recalled in Appendix A, all the lower levels can be obtained via optimization
of the level 2.5 in the presence of constraints, in particular the level 2
concerning the distribution $P^{[2]}_T \left[ {\mathring P} (.)  \right] $ 
of the empirical probability ${\mathring P}(\sigma_.)$ alone,
that can be obtained from the integration of the 
joint distribution $P^{[2.5]}_T \left[ {\mathring P} (.) ; {\mathring W}(.\vert.) \right] $ 
of Eq. \ref{level2.5chain}
over the empirical kernel ${\mathring W}(\sigma_. \vert  {\tilde \sigma}_.) $ as follows
\begin{eqnarray}
&& P^{[2]}_T \left[ {\mathring P} (.)  \right]  \equiv \int d {\mathring W}(.\vert.)
P^{[2.5]}_T \left[ {\mathring P} (.) ; {\mathring W}(.\vert.) \right] 
\nonumber \\
&& \opsimeq_{T \to + \infty}
   \delta \left( \sum_{\sigma_.} {\mathring P}(\sigma_.)  - 1 \right) 
   \int d {\mathring W}(.\vert.)
  \left[ \prod_{ {\tilde \sigma}_.} \delta \left(  \sum_{\sigma_.}  {\mathring W}(\sigma_. \vert  {\tilde \sigma}_.) - 1 \right) \right]
 \left[\prod_{ \sigma_.} \delta \left(  \sum_{{\tilde \sigma}_.}   {\mathring W}(\sigma_. \vert  {\tilde \sigma}_.) {\mathring P}({\tilde \sigma}_.)- {\mathring P}(\sigma_.) \right)  \right] 
 e^{\displaystyle -T  I_{2.5}\left[ {\mathring P} (.) ; {\mathring W}(.\vert.)\right] }
 \nonumber \\
&& \opsimeq_{T \to + \infty}
   \delta \left( \sum_{\sigma_.} {\mathring P}(\sigma_.)  - 1 \right) 
 e^{\displaystyle -T  I_2\left[ {\mathring P} (.) \right] }
\label{level2chain}
\end{eqnarray}
So the rate function $I_2\left[ {\mathring P} (.)\right] $ at level 2 corresponds to the optimization
of the rate function  $I_{2.5}\left[ {\mathring P} (.) ; {\mathring W}(.\vert.)\right]$
over the empirical kernel ${\mathring W}(\sigma_. \vert  {\tilde \sigma}_.) $
satisfying the constitutive constraints, but for an arbitrary Markov chain,
the solution of this constrained optimization
cannot be written as an explicit function of the empirical probability ${\mathring P}(\sigma_.)$.

In the next section, we return to the special case of the backward stochastic dynamics of the doubling map in the binary variables in order to
obtain what levels of large deviations can be written explicitly.


\subsection{ Application to the backward stochastic dynamics of the doubling map in the binary variables } 

Let us apply Eq. \ref{level2.5chain} to the special case of the backward kernel $W_B(\sigma_. \vert  {\tilde \sigma }_. ) $ of Eq. \ref{Wdoublingsigmabackward} in the binary variables associated to the the doubling map.
In order for the rate function $ I_{2.5} \left[ {\mathring P} (.) ; {\mathring W}(.\vert.)\right] $ of Eq. \ref{rate2.5chain} to remain finite,
the empirical backward kernel ${\mathring W}_B(\sigma_. \vert  {\tilde \sigma }_. ) $ has to be of the form
\begin{eqnarray}
{\mathring W}_B(\sigma_. \vert  {\tilde \sigma }_. )  
&&  =\left[ \delta_{ \sigma _1,0} {\mathring p}_{B-}({\tilde \sigma }_. )
+ \delta_{ \sigma _1,1} {\mathring p}_{B+}({\tilde \sigma }_. )\right] \prod_{l=2}^{+\infty} \delta_{\sigma_l,{\tilde \sigma }_{l-1}}
  \label{Wempibackbinary}
\end{eqnarray}
where the normalization condition of the
empirical kernel ${\mathring W}(\sigma_. \vert  {\tilde \sigma}_.) $
over $\sigma_. $ for any ${\tilde \sigma}_. $ reads
\begin{eqnarray}
1= \sum_{\sigma_.}  {\mathring W}_B(\sigma_. \vert  {\tilde \sigma}_.)
  = {\mathring p}_{B-}({\tilde \sigma }_. )
+  {\mathring p}_{B+}({\tilde \sigma }_. )
  \label{Wempibackbinaryii}
\end{eqnarray}
The condition (iii) imposing that the empirical density 
${\mathring P}_B(\sigma_.)$ is steady with respect to the 
dynamics governed by
the empirical kernel ${\mathring W}_B(\sigma_. \vert  {\tilde \sigma}_.) $ 
\begin{eqnarray}
{\mathring P}_B(\sigma_1,\sigma_2,...)
=\sum_{{\tilde \sigma}_.}   {\mathring W}_B(\sigma_. \vert  {\tilde \sigma}_.) {\mathring P}_B({\tilde \sigma}_.)
= \left[ \delta_{ \sigma _1,0} {\mathring p}_{B-}(\sigma_2,\sigma_3,...)
+ \delta_{ \sigma _1,1} {\mathring p}_{B+}(\sigma_2,\sigma_3,...)\right]  {\mathring P}_B(\sigma_2,\sigma_3,...)
  \label{Wempibackbinaryiii}
\end{eqnarray}
can be used to compute the two complementary empirical probabilities $ {\mathring p}_{B\pm}({\tilde \sigma }_. )$ that parametrize the empirical kernel of Eq. \ref{Wempibackbinary}
in terms of the empirical probability ${\mathring P}(.) $
\begin{eqnarray}
 {\mathring p}_{B-}(\sigma_2,\sigma_3,...) && = \frac{ {\mathring P}_B(0,\sigma_2,\sigma_3,...) }
 { {\mathring P}_B(\sigma_2,\sigma_3,...) }
\nonumber \\
 {\mathring p}_{B+}(\sigma_2,\sigma_3,...) && = \frac{ {\mathring P}_B(1,\sigma_2,\sigma_3,...) }
{  {\mathring P}_B(\sigma_2,\sigma_3,...) }
  \label{Wempibackbinaryiiiinvert}
\end{eqnarray}
so that the constraint of Eq. \ref{Wempibackbinaryii}
becomes
\begin{eqnarray}
1  = {\mathring p}_{B-}({\tilde \sigma }_. )
+  {\mathring p}_{B+}({\tilde \sigma }_. )
=  \frac{ {\mathring P}_B(0,\sigma_2,\sigma_3,...) + {\mathring P}_B(1,\sigma_2,\sigma_3,...) }
{  {\mathring P}_B(\sigma_2,\sigma_3,...) }
  \label{Wempibackbinaryiidensity}
\end{eqnarray}
in terms of the empirical probability.

As a consequence, in our present very specific case, 
the empirical backward kernel ${\mathring W}_B(\sigma_. \vert  {\tilde \sigma }_. ) $ 
of Eq. \ref{Wempibackbinary} parametrized 
by the two complementary empirical probabilities $ {\mathring p}_{B\pm}({\tilde \sigma }_. )$
can be rewritten in terms of the empirical empirical probability ${\mathring P}(.) $
using Eq. \ref{Wempibackbinaryiiiinvert}.
So
the application of the level 2.5 of Eq. \ref{level2.5chain}
for arbitrary Markov chains 
actually gives, for our present very specific case, 
the following explicit large deviation form at level 2 for
the probability of the empirical probability ${\mathring P}_B (.) $ alone
\begin{eqnarray}
P^{[2]}_T \left[ {\mathring P}_B (.)  \right] 
&& \opsimeq_{T \to + \infty}
   \delta \left( \sum_{\sigma_.} {\mathring P}_B(\sigma_.)  - 1 \right) 
 \left[\prod_{ \sigma_2,\sigma_3,...} 
 \delta \left( {\mathring P}_B(0,\sigma_2,\sigma_3,...) + {\mathring P}_B(1,\sigma_2,\sigma_3,...)  - {\mathring P}_B(\sigma_2,\sigma_3,) \right)  \right] \ \nonumber \\
 && e^{\displaystyle -T  I_2\left[ {\mathring P}_B (.) \right] }
\label{level2backwardsigma}
\end{eqnarray}
where the constraint
of Eq. \ref{Wempibackbinaryiidensity} 
has to be added to the usual normalization constraint
of the empirical probability,
while the rate function $I_{2}\left[ {\mathring P}_B (.) \right] $ at level 2 
obtained from Eq. \ref{rate2.5chain} reads 
using Eqs \ref{Wempibackbinary}
and \ref{Wempibackbinaryiiiinvert}
\begin{eqnarray}
&&  I_2 \left[ {\mathring P}_B (.) \right]
\nonumber \\
&&  =   \sum_{\sigma_.}
  \left[ \delta_{ \sigma _1,0} \frac{ {\mathring P}_B(0,\sigma_2,\sigma_3,...) }
 { {\mathring P}_B(\sigma_2,\sigma_3,...) }
+ \delta_{ \sigma _1,1} \frac{ {\mathring P}_B(1,\sigma_2,\sigma_3,...) }
{  {\mathring P}_B(\sigma_2,\sigma_3,...) }\right] 
    {\mathring P}_B(\sigma_2,\sigma_3,...)
  \ln \left[ \delta_{ \sigma _1,0} 2 \frac{ {\mathring P}_B(0,\sigma_2,\sigma_3,...) }
 { {\mathring P}_B(\sigma_2,\sigma_3,...) }
+ \delta_{ \sigma _1,1} 2 \frac{ {\mathring P}_B(1,\sigma_2,\sigma_3,...) }
{  {\mathring P}_B(\sigma_2,\sigma_3,...) }\right] 
\ \ \ 
\nonumber \\
&& = \sum_{ \sigma_2,\sigma_3,...} 
\left( 
 {\mathring P}_B(0,\sigma_2,\sigma_3,...) 
  \ln \left[  2 \frac{ {\mathring P}_B(0,\sigma_2,\sigma_3,...) }
 { {\mathring P}(\sigma_2,\sigma_3,...) }\right] 
+\left[   {\mathring P}_B(1,\sigma_2,\sigma_3,...) \right] 
  \ln \left[  2 \frac{ {\mathring P}_B(1,\sigma_2,\sigma_3,...) }
{  {\mathring P}_B(\sigma_2,\sigma_3,...) }\right] 
\right)
\nonumber \\
&& = \sum_{ \sigma_1,\sigma_2,\sigma_3,...} 
 {\mathring P}_B(\sigma_1,\sigma_2,\sigma_3,...) 
  \ln \left[  2 \frac{ {\mathring P}_B(\sigma_1,\sigma_2,\sigma_3,...) }
 { {\mathring P}_B(\sigma_2,\sigma_3,...) }\right] 
\label{rate2backwardsigma}
\end{eqnarray}

In summary, while for arbitrary Markov chains only the level 2.5 
for the joint probability $P^{[2.5]}_T \left[ {\mathring P} (.) ; {\mathring W}(.\vert.) \right]  $ of Eq. \ref{level2.5chain} 
is always explicit,
we have obtained that for the specific case of the backward stochastic Markov chains associated to
the forward deterministic doubling map, the empirical kernel ${\mathring W}(.\vert.) $  is actually determined by the empirical probability $ {\mathring P} (.) $ so
that the explicit level 2.5 produces 
an explicit level 2 concerning the distribution of $P^{[2]}_T \left[ {\mathring P}_B (.)  \right]  $
of the empirical probability ${\mathring P}_B (.) $ alone.
This level reduction from 2.5 towards 2 is expected to be general for chaotic non-invertible
maps, and can be considered as the counterpart of the reduction of Eq. \ref{observablemap} 
concerning trajectory observables.


\section{ Conclusions }

\label{sec_conclusion}

In this paper, we have revisited the recent studies \cite{naftali,spain}
concerning the large deviations properties of trajectory observables for chaotic non-invertible deterministic maps in order to analyze in detail the similarities and the differences with the case of stochastic Markov chains recalled in Appendix A. To be concrete, we have focused on the simplest example of the doubling map on the real-space interval $x \in [0,1[$ that can be also analyzed via the decomposition $x= \sum_{l=1}^{+\infty} \frac{\sigma_l}{2^l} $ into binary coefficients $\sigma_l=0,1$. Since the large deviations properties of trajectory observables can be studied either via deformations of the forward deterministic dynamics or via deformations of the backward stochastic dynamics, either in real space or in binary space, we have discussed the advantages and the drawbacks of these various perspectives, and we have illustrated them with the explicit example of the simplest trajectory observable $\omega(x)=x$ 
already considered in \cite{naftali,spain}.
Let us now summarize the main conclusions 
and discuss what happens in other deterministic chaotic systems:

(1) The $k$-deformation of the forward deterministic dynamics can be used to obtain the 
deformed dominant eigenvalue $\lambda_0^{[k]}$ governing the scaled cumulants of trajectory observables,
with its corresponding left and right eigenvectors:

(1a) We have stressed 
that the dominant deformed left eigenvector $l_0^{[k]}(.) $ is singular in real space,
since it involves at first order in perturbation theory in $k$ the non-dominant left eigenvector $l_p^{[0]}(.) $
of the unperturbed forward dynamics, that are expected to be always singular in real space
as a consequence of the sensitivity with respect to initial conditions of any deterministic chaotic dynamics.

(1b) For the doubling map, we have stressed that the corresponding 
dominant deformed right eigenvector $r_0^{[k]}(.) $ is regular in real space, 
as are the non-dominant right eigenvectors $r_p^{[0]}(.) $ of the unperturbed forward dynamics.
This property should hold for other non-invertible maps, whose backward dynamics
is stochastic at each time step, since this stochastic noise
 will erase the information on the initial condition and will separate trajectories independently
 of the details of their initial conditions.
However for invertible maps whose backward dynamics is deterministic chaotic,
both the dominant deformed right eigenvector $r_0^{[k]}(.) $ 
and the non-dominant undeformed right eigenvectors $r_p^{[0]}(.) $ 
will be be singular in real space.
Note that for deterministic chaotic systems with strange attractors, even the 
dominant undeformed right eigenvector $r_0^{[0]}(.) = \rho_*(.)$ is already singular in real space.

(2) For chaotic non-invertible maps, we have stressed two important advantages
of the backward stochastic dynamics in the symbolic space: 

(2a) For each trajectory observable, an appropriate Doob canonical conditioned dynamics 
different from the initial dynamics can be constructed only in the backward stochastic perspective 
where the reweighting of existing transitions is possible, and not in the forward deterministic perspective. 
The corresponding conditioned steady state $ \rho^{C[k]}_*(.) =l_0^{[k]}(.) r_0^{[k]}(.) $ 
 given by the product of the dominant deformed left and right eigenvectors
 is always singular in real space, as a consequence of the singular 
 character of the left eigenvector $l_0^{[k]}(.)$ 
 any deterministic chaotic dynamics as discussed in 1a) above.
 As a consequence, it is more appropriate to analyze this conditioned dynamics in symbolic space,
 as shown on an explicit example for the doubling map where the symbolic space corresponds
  to infinite binary strings.

(2b)  The backward stochastic dynamics in symbolic space is also the most appropriate framework to 
analyze large deviations at higher levels beyond the level of trajectory observables.
We have explained that the application of the explicit large deviations at level 2.5 for arbitrary Markov chains
to the specific case of the stochastic backward dynamics associated to forward chaotic deterministic
non-invertible maps
produces explicit large deviations at level 2 for the empirical density alone.
In accordance with this level reduction from the level 2.5 towards the level 2,
the general trajectory observables of Markov chains that involve both
the empirical density and the empirical flows can be rewritten in terms of the empirical density alone,
and thus reduce to the trajectory observables of the so-called Level 1, 
that can be obtained from the Level 2 via contraction.

Note that for chaotic invertible maps, whose backward dynamics is not stochastic but deterministic,
one looses these advantages summarized in (2a) and (2b), but the symbolic dynamics 
is even more necessary to analyze the large deviations of trajectory observables
via deformed generators, since both the dominant deformed right and left eigenvectors 
are singular in real space as discussed in points (1a) and (1b) above.


\appendix

\section{ Reminder on large deviations at various levels for stochastic Markov chains }

\label{app_markov}

In this Appendix, we recall the large deviations properties at various levels
for stochastic Markov chains, in order to compare with the main text concerning chaotic maps.


\subsection{ Analysis of the Markov chain dynamics via the spectral decomposition of its kernel $w({\tilde x} \vert x)$  }

\label{sec_markovchain}

The evolution of the probability density $\rho_t(  x)  $ to be at position $ x$ at time $t$
\begin{eqnarray}
\rho_{t+1}(  {\tilde x}) =  \int d  x  \ w({\tilde x} \vert x)  \rho_t(  x) 
\label{markovchainc}
\end{eqnarray}
is governed by the Markov-Chain kernel $w( x,  {\tilde x})  $ satisfying the normalization
\begin{eqnarray}
   \int d {\tilde x}  \ w({\tilde x} \vert x)  =1 \ \ \ \text{ for any }  x
\label{markovnorma}
\end{eqnarray}
We focus on cases where the dynamics of Eq. \ref{markovchainc}
converges towards some normalizable steady-state $\rho_*(  .) $ satisfying
the time-independent version of Eq. \ref{markovchainc}
\begin{eqnarray}
\rho_*(  {\tilde x}) =  \int d  x  \ w({\tilde x} \vert x)  \rho_*(  x) 
\label{markovchaincsteady}
\end{eqnarray}

The convergence towards this steady state can be analyzed
via the spectral decomposition of the kernel $w({\tilde x} \vert x)$
\begin{eqnarray}
w( {\tilde x} \vert  x) =  \langle  {\tilde x} \vert \bigg(   \sum_{p=0}^{+\infty} \lambda_p \vert r_p \rangle \langle l_p \vert \bigg) \vert x \rangle
= \sum_{p=0}^{+\infty} \lambda_p r_p({\tilde x}) l_p(x)
\label{Wspectral}
\end{eqnarray}
in terms of its eigenvalues $\lambda_p$
with their right eigenvectors $r_p(  .) $ 
and left eigenvectors $l_p(  .) $ satisfying
\begin{eqnarray}
\lambda_p  r_p(  {\tilde x}) && =  \int d  x  \ w({\tilde x} \vert x)  r_p(  x)
\nonumber \\
\lambda_p  l_p(  x) &&  =  \int d {\tilde x}  l_p(  {\tilde x}) w({\tilde x} \vert x)  
\label{eigenW}
\end{eqnarray}
with the orthonormalization
\begin{eqnarray}
\delta_{p,p'}= \langle l_p \vert r_{p'} \rangle = \int  dx  l_p(  x)r_{p'}(  x)
\label{orthonorma}
\end{eqnarray}
Eqs \ref{markovchainc} and \ref{markovnorma} mean that the highest eigenvalue is $\lambda_{p=0}=1$
with the corresponding positive right and left eigenvectors
\begin{eqnarray}
 r_0(  {\tilde x}) && =  \rho_*(  {\tilde x}) 
\nonumber \\
 l_0(  x) &&  =  1  
\label{eigenW1}
\end{eqnarray}

Then the density $\rho_t(.) $ at time $t$ can be written in terms of the initial density $\rho_{t=0}(.) $
using the spectral decomposition of Eq. \ref{Wspectral}
\begin{eqnarray}
\rho_t({\tilde x}) && =  \sum_{p=0}^{+\infty} \lambda_p^t r_p(  {\tilde x}) \int  dx_0 l_p(x_0) \rho_0(x_0)
\nonumber \\
&& = \rho_*(  {\tilde x}) +  \sum_{p=1}^{+\infty} \lambda_p^t r_p(  {\tilde x}) \int  dx_0 l_p(x_0) \rho_0(x_0)
\label{markovspectral}
\end{eqnarray}


\subsection{ Large deviations of trajectory observables via appropriate deformations of Markov kernel}

\label{subsec_trajviadeform}

Let us now consider an observable ${\cal O}^{Traj} [ x_0,x_1,...,x_T ]$ of the trajectory $(x_0,x_1,...,x_T)$
that can be parametrized by a function $O( {\tilde x} ,  x)$
\begin{eqnarray}
{\cal O}^{Traj}\left[ x_0,x_1,...,x_T \right]  
=   \frac{1}{T  } \sum_{t=0}^{T-1} O ( x_{t+1},  x_t)
\label{observablerho2}
\end{eqnarray}
i.e. it is a time-average of the function $O ( x_{t+1},  x_t) $ involving two consecutive positions.


\subsubsection{ Generating function $Z_T^{[k]} (x_T \vert x_0) $ of the trajectory observable via
via the appropriate deformed kernel $w^{[k]}({\tilde x} \vert x)$  }

The standard method to analyze the statistics of the trajectory observable of Eq. \ref{observablerho2}
is based on the generating function of ${\cal O}^{Traj} $
over the trajectories starting at $x(t=0)=x_0 $ and ending at $x(t=T)=x_T $
\begin{eqnarray}
Z_T^{[k]} (x_T \vert x_0) \equiv  \langle e^{ \displaystyle k T {\cal O}^{Traj} } \rangle
&& = \int dx_1 \int dx_2 ... \int dx_{T-1} 
 \prod_{t=0}^{T-1} \left[ w(x_{t+1},x_t ) e^{ k O ( x_{t+1},  x_t)} \right]
 \nonumber \\
 && \equiv  \int dx_1 \int dx_2 ... \int dx_{T-1}
  \prod_{t=0}^{T-1} w^{[k]}(x_{t+1},x_t ) \equiv \langle x_T \vert  \left(w^{[k]}\right)^T \vert x_0\rangle 
\label{genektraj}
\end{eqnarray}
that is governed by the $k$-deformed kernel
\begin{eqnarray}
 w^{[k]}({\tilde x} \vert x) \equiv w({\tilde x} \vert x) e^{ k O ( {\tilde x},x)}
\label{Wkdeformed}
\end{eqnarray}


\subsubsection{ Eigenvalue problem for $w^{[k]}({\tilde x} \vert x)$ 
to obtain the scaled-cumulant generating function 
of the trajectory observable }

For large time $T \to + \infty$, the generating function of Eq. \ref{genektraj}
is governed 
\begin{eqnarray}
Z_T^{[k]} (x_T \vert x_0) =\langle x_T \vert  \left(w^{[k]}\right)^T \vert x_0\rangle  
\opsimeq_{T \to +\infty}  [\lambda_0^{[k]}]^T r_0^{[k]}(x_T) l_0^{[k]}(x_0) 
\label{Wktiltprop}
\end{eqnarray}
by the highest eigenvalue $\lambda_0^{[k]}$ associated to the positive right eigenvector 
$r_0^{[k]}(  {\tilde x}) \geq 0 $ and to the positive left eigenvector $l_0^{[k]}(  {\tilde x}) \geq 0 $
satisfying
\begin{eqnarray}
\lambda_0^{[k]} \ r_0^{[k]}(  {\tilde x}) && =  \int d  x  \ w^{[k]}({\tilde x} \vert x)  r_0^{[k]}(  x) 
= \int d  x  \ w({\tilde x} \vert x) e^{ k O ( {\tilde x},x)}  r_0^{[k]}(  x)
\nonumber \\
\lambda_0^{[k]}  l_0^{[k]}(  x) &&  =  \int d {\tilde x}  l_0^{[k]}(  {\tilde x}) w^{[k]}({\tilde x} \vert x) 
=  \int d {\tilde x}  l_0^{[k]}(  {\tilde x}) w({\tilde x} \vert x) e^{ k O ( {\tilde x},x)} 
\label{eigenWk}
\end{eqnarray}
with the normalization
\begin{eqnarray}
 \int dx  l_0^{[k]}(x) r_0^{[k]}(x) =1
 \label{Wktiltnorma}
\end{eqnarray}
while for $k=0$ one recovers the eigenvectors of Eq \ref{eigenW1} associated the highest eigenvalue $\lambda^{[k=0]}_0=1$ with its eigenvectors of Eq. \ref{eigenW1}.

From the generating function $ Z_T^{[k]} (x_T \vert x_0)$ that will grow or decay in time as $[\lambda_0^{[k]}]^T $, there are two simple ways to construct processes
that conserve the probability as we now recall.


\subsubsection{ Construction of the normalized density $\rho^{[k]End}_T(x \vert x_0) 
= \frac{Z_T^{[k]} (x \vert x_0)}{ \int dy Z_T^{[k]} (y \vert x_0)}$ for the end-point $x$ }

The simplest way to construct a normalized probability out of the generation function $Z_T^{[k]} (x \vert x_0) $
is to consider the probability of the end-point
\begin{eqnarray}
\rho^{[k]End}_T(x \vert x_0) 
\equiv \frac{Z_T^{[k]} (x \vert x_0)}{ \int dy Z_T^{[k]} (y \vert x_0)}
 \label{probaEndk}
\end{eqnarray}

The asymptotic behavior of Eq. \ref{Wktiltprop} for the generating function yields
that the probability of the end point $x$ 
\begin{eqnarray}
\rho^{[k]End}_T(x \vert x_0) \opsimeq_{T \to +\infty}
\frac{r_0^{[k]}(x)}{ \int dy r_0^{[k]}(y)} \equiv \rho^{[k]End}_*(x)
 \label{probaEndksteady}
\end{eqnarray}
converges, independently of the initial point $x_0$
 towards the distribution $\rho^{[k]End}_*(x) $ determined by the appropriate normalization
 right eigenvector $r_0^{[k]}(.) $.


\subsubsection{ Construction of the conditioned kernel $w^{C[k]}({\tilde x} \vert x) $ via the Doob's transform involving the left eigenvector $l_0^{[k]}(.) $ }

Another interesting way to construct a normalized density 
out of the generation function $Z_T^{[k]} (x \vert x_0) $
is via the Doob's conditioning similarity transform involving the left eigenvector $l_0^{[k]}(.) $ of Eq. 
Eq. \ref{eigenWk}
\begin{eqnarray}
\rho^{C[k]}_T(x \vert x_0) 
&& \equiv  \frac{1}{ [\lambda_0^{[k]}]^T} \  l_0^{[k]}(x)  Z_T^{[k]} (x \vert x_0)  \frac{1}{ l_0^{[k]}(x_0)  }
=  \frac{1}{[\lambda_0^{[k]}]^T} \  l_0^{[k]}(x)  \langle x \vert  \left(w^{[k]}\right)^T \vert x_0\rangle  \frac{1}{ l_0^{[k]}(x_0)  }
\nonumber \\
&& \equiv \langle x \vert  \left(w^{C[k]}\right)^T \vert x_0\rangle
  \label{ProbaDoobk}
\end{eqnarray}
where the conditioned kernel
\begin{eqnarray}
w^{C[k]}({\tilde x} \vert x) 
&& \equiv  \frac{1}{\lambda_0^{[k]}} \  l_0^{[k]}({\tilde x}) w^{[k]}({\tilde x} \vert x)   \frac{1}{ l_0^{[k]}(x)  }
\nonumber \\
&& = \frac{l_0^{[k]}({\tilde x}) w^{[k]}({\tilde x} \vert x)}{\int d  z  l_0^{[k]}(  z) w^{[k]}(z \vert x)  }
= \frac{l_0^{[k]}({\tilde x}) w({\tilde x} \vert x) e^{ k O ( {\tilde x},x)}}{\int d  z  l_0^{[k]}(  z) w(z \vert x) e^{ k O ( z,x)} }
  \label{WDoobk}
\end{eqnarray}
has been rewritten using the eigenvalue Eq. \ref{eigenWk} for the left eigenvector $  l_0^{[k]}(x)$ 
in the denominator to obtain the last expression that makes obvious the normalization over $\tilde x$ for any $x$. The physical meaning of Eq. \ref{WDoobk}
is that the transitions $ x \to {\tilde x}$ that were possible with the undeformed kernel
$w({\tilde x} \vert x) \ne 0$ are reweighted, 
while the transitions $ x \to {\tilde x}$ that were not possible with the undeformed kernel
$w({\tilde x} \vert x) =0 $ are still impossible.

Plugging the asymptotic behavior of Eq. \ref{Wktiltprop} for the generating function
into Eq. \ref{ProbaDoobk} yields that for large $T$, the density $\rho^{C[k]}_T(x \vert x_0) $
\begin{eqnarray}
\rho^{C[k]}_T(x \vert x_0) 
&& \equiv  \frac{1}{ [\lambda_0^{[k]}]^T} \  l_0^{[k]}(x)  Z_T^{[k]} (x \vert x_0)  \frac{1}{ l_0^{[k]}(x_0)  }
\opsimeq_{T \to +\infty}  l_0^{[k]}(x)  r_0^{[k]}(x) \equiv \rho^{C[k]}_*( x) 
  \label{ProbaDoobksteady}
\end{eqnarray}
converges, independently of the initial point $x_0$
 towards the conditioned steady distribution $\rho^{C[k]}_*( x) $ 
 given by the product of the left eigenvector $l_0^{[k]}(.)  $ and of the right eigenvector $r_0^{[k]}(.) $,
 whose normalization over $x$ is ensured by Eq. \ref{Wktiltnorma}.

In order to better understand the different physical meaning with respect to the end-point distribution of Eqs \ref{probaEndk}
and \ref{probaEndksteady}, it is useful to consider
the density of position $x$ at some interior time $0 \ll t \ll T$
using the asymptotic expression of Eq. \ref{Wktiltprop} for the three 
generating functions on the time-intervals $[0,t]$, $[t,T]$ and $[0,T]$
\begin{eqnarray}
\rho_t^{[k]Interior}(x) && \equiv 
\frac{ Z_{T-t}^{[k]} (x_T \vert x)  Z_t^{[k]} (x \vert x_0) }
{ Z_T^{[k]} (x_T \vert x_0) } 
\nonumber \\
&&\opsimeq_{T \to +\infty}  \frac{[\lambda_0^{[k]}]^{T-t} r_0^{[k]}(x_T) l_0^{[k]}(x) 
[\lambda_0^{[k]}]^t r_0^{[k]}(x) l_0^{[k]}(x_0) }
{[\lambda_0^{[k]}]^T r_0^{[k]}(x_T) l_0^{[k]}(x_0) }
=  l_0^{[k]}(x) r_0^{[k]}(x) \equiv \rho^{C[k]}_*( x) 
  \label{rhosteadyDoob}
\end{eqnarray}
So the conditioned steady state $\rho^{C[k]}_*( x)  $
corresponds to the distribution of the position $x$ seen at interior times $0 \ll t \ll T$
independently of the initial position $x_0$ and of the final position $x_T$.


\subsubsection{ Physical meaning of the largest deformed eigenvalue $\lambda_0^{[k]}$   }

\label{subsec_leveltraj}

The dominant behavior of Eq. \ref{Wktiltprop} that is independent of the two positions $(x_0;x_T)$
\begin{eqnarray}
Z_T^{[k]} 
\opsimeq_{T \to +\infty}  [\lambda_0^{[k]}]^T \equiv e^{ \displaystyle T  c(k) }
\ \ \ {\rm with } \ \ c(k) \equiv   \ln \lambda_0^{[k]} 
\label{Wktiltpropexp}
\end{eqnarray}
means that $ c(k) \equiv   \ln \lambda_0^{[k]} $ represents the generating function 
of the scaled-cumulants of the trajectory observable ${\cal O}^{Traj}\left[ x_0,x_1,...,x_T \right] $ 
of Eq. \ref{observablerho2} (see the review \cite{review_touchette} for more details).

The link with the rate function $I({\cal O})$ governing the large deviations for large $T$
of the probability $P_T({\cal O}) $
to see the value ${\cal O}$ for the trajectory observable ${\cal O}^{Traj}\left[ x_0,x_1,...,x_T \right] $ 
of Eq. \ref{observablerho2}
\begin{eqnarray}
P_T({\cal O})
\opsimeq_{T \to +\infty}   e^{ \displaystyle - T  I({\cal O}) }
\label{ldtraj}
\end{eqnarray}
can be understood via the computation of the generating function of Eqs \ref{genektraj} and \ref{Wktiltpropexp}
via the saddle-point evaluation for large $T$
\begin{eqnarray}
Z_T^{[k]} && \equiv  \langle e^{ \displaystyle k T {\cal O} } \rangle
= \int d {\cal O} P_T({\cal O}) e^{ \displaystyle k T {\cal O} }
 \nonumber \\
 &&  \opsimeq_{T \to +\infty} \int d {\cal O}  e^{ \displaystyle  T \left[ k {\cal O} -  I({\cal O})\right] }
 \opsimeq_{T \to +\infty}   e^{ \displaystyle T  c(k) }
\label{genektrajsaddle}
\end{eqnarray}
 that leads to the Legendre transformation
\begin{eqnarray}
 k {\cal O} -  I({\cal O}) && =  c(k) 
 \nonumber \\
 k -  I'({\cal O}) && =0
\label{legendre}
\end{eqnarray}
with its inverse
\begin{eqnarray}
I({\cal O})&& = k {\cal O} -    c(k) 
 \nonumber \\
0 && ={\cal O} -    c'(k) 
\label{legendreinverse}
\end{eqnarray}

In summary, the largest deformed eigenvalue $\lambda_0^{[k]}$
gives directly the scaled cumulant generating function $c(k)= \left[  \ln \lambda_0^{[k]} \right]$,
whose Legendre transform is the rate function $ I({\cal O}) $
governing the large deviations of Eq. \ref{ldtraj} for the trajectory observable.


\subsubsection{ Perturbation theory in $k$ of the eigenvalue $\lambda_0^{[k]}$ and for the eigenvectors  }

\label{subsec_permarkov}

When it is not possible to solve exactly the eigenvalue equations of Eq. \ref{eigenWk}
for the deformed kernel $w^{[k]}({\tilde x} \vert x) $,
it is useful to consider the perturbation theory with respect to the undeformed kernel 
$w^{[k=0]}({\tilde x} \vert x) =w({\tilde x} \vert x)$
where the eigenvalue $\lambda_0^{[k=0]} =1 $ is known with its eigenvectors of Eq. \ref{eigenW1}. 

The series expansion in $k$ for the deformed kernel of Eq. \ref{Wkdeformed}
\begin{eqnarray}
 w^{[k]}({\tilde x} \vert x) = w({\tilde x} \vert x) e^{ k O ( {\tilde x},x)}
 = w({\tilde x} \vert x)  + k  w^{(1)}({\tilde x} \vert x)  + k^2  w^{(2)}({\tilde x} \vert x) +o(k)
\label{Wkper}
\end{eqnarray}
involves the first-order and second-order corrections
\begin{eqnarray}
 w^{(1)}({\tilde x} \vert x) && = w({\tilde x} \vert x)  O ( {\tilde x},x)
 \nonumber \\
 w^{(2)}({\tilde x} \vert x)  && = w({\tilde x} \vert x) \frac{ O^2 ( {\tilde x},x)} {2} 
\label{Wkper12}
\end{eqnarray}
The perturbation theory for the highest eigenvalue
\begin{eqnarray}
\lambda_0^{[k]} = 1  + k  \lambda^{(1)}  + k^2 \lambda^{(2)}   +o(k)
\label{lambdaper}
\end{eqnarray}
is very similar to the standard theory of quantum mechanics with the following conclusions.

The fist-order correction $\lambda^{(1)} $ only involves the first-order perturbation $w^{(1)}({\tilde x} \vert x) $
and the unperturbed eigenvectors of Eq. \ref{eigenW1}
\begin{eqnarray}
  \lambda^{(1)}  = \langle  l_0 \vert w^{(1)} \vert r_0   \rangle 
&&  = \int d {\tilde x} \int dx l_0 ({\tilde x}) w^{(1)}({\tilde x} \vert x)  r_0 (x)
  = \int dx \int d {\tilde x}  w({\tilde x} \vert x)  O ( {\tilde x},x)  \rho_* (x) 
\label{lambdaper1}
\end{eqnarray}
and coincides with the steady value of the trajectory observable that can be computed 
in terms of the steady state $ \rho_* (x)  $.

The first-order corrections for the right and left eigenvectors with respect to their unperturbed values of Eqs \ref{eigenW1}
\begin{eqnarray}
 r_0^{[k]}(x) && =  r_0 (x) + k  r^{(1)}_0 (x) + o(k)  = \rho_*(x)+ k  r^{(1)}_0 (x) + o(k) 
 \nonumber \\
 l_0^{[k]}(  x)   && = l_0 (x)+k  l^{(1)}_0 (x) + o(k)=  1 +k  l^{(1)}_0 (x) + o(k)
\label{eigenvecper}
\end{eqnarray}
can be written
\begin{eqnarray}
  \vert r^{(1)}_0 \rangle && = G w^{(1)} \vert r_0  \rangle
    \nonumber \\
 \langle l^{(1)}_0 \vert && =\langle l_0 \vert w^{(1)} G
 \label{eigenvecper1ket}
\end{eqnarray}
in terms of the Green function 
\begin{eqnarray}
G   \equiv \left(\mathbb{1}- \vert r_0 \rangle \langle  l_0 \vert\right) 
\frac{1}{\mathbb{1} - w }  \left(\mathbb{1}- \vert r_0 \rangle \langle  l_0 \vert\right) 
 =\sum_{p=1}^{+\infty} \frac{ \vert r_p \rangle \langle  l_p \vert }{1-\lambda_p}
\label{green}
\end{eqnarray}
The first expression means that the Green function $G$ is the inverse of the operator $(\mathbb{1} - w  )$
 in the subspace $\left(\mathbb{1}- \vert r_0 \rangle \langle  l_0 \vert\right) $ orthogonal 
 to the subspace $\vert r_0 \rangle \langle  l_0 \vert $ associated to the unity eigenvalue $\lambda_0=1$
 The second expression involves all the unperturbed other eigenvalues
$\lambda_{p \geq 1}$ with their right and left eigenvectors
that were discussed for the spectral decomposition of Eq. \ref{markovspectral}.

So the first-order corrections of Eqs \ref{eigenvecper1ket} 
\begin{eqnarray}
   r^{(1)}_0 ({\tilde x}) && = \int dz \int dx G({\tilde x},z) w^{(1)}(z \vert x) r_0 (x) 
   =\sum_{p=1}^{+\infty} r_p ({\tilde x})  
   \left[ \frac{  1  }{1-\lambda_p}  \int dx
     \int dz l_p (z) w(z \vert x)  O (z,x)   \rho_*(x)  \right]
    \nonumber \\
  l^{(1)}_0 (z)  && = \int dx \int d {\tilde x} l_0 (x) w^{(1)}(x \vert {\tilde x}) G({\tilde x},z) 
  =  \sum_{p=1}^{+\infty} l_p (z) \left[ \frac{ 1 }{1-\lambda_p}  
  \int d {\tilde x}  \int dx l_0 (x) w(x \vert {\tilde x}) O(x,{\tilde x})  r_p ({\tilde x}) \right]
\label{eigenvecper1}
\end{eqnarray}
are decomposed onto all the other unperturbed right eigenvectors $r_p ({\tilde x}) $ 
and on all the other unperturbed left eigenvectors $ l_p (z) $ respectively.

The first-order correction for the conditioned density of Eq. \ref{rhosteadyDoob}
\begin{eqnarray}
\rho^{C[k]}_*( {\tilde x}) =  l_0^{[k]}({\tilde x}) r_0^{[k]}({\tilde x}) = \rho_*({\tilde x}) + k \rho^{C(1)}_*({\tilde x}) + o(k)
  \label{rhosteadyDoobper}
\end{eqnarray}
is given by
\begin{eqnarray}
 \rho^{C(1)}_*({\tilde x}) = l_0({\tilde x}) r_0^{(1)}({\tilde x})+l_0^{(1)}({\tilde x}) r_0({\tilde x})
 =  r_0^{(1)}({\tilde x})+l_0^{(1)}({\tilde x}) \rho_*({\tilde x})
  \label{rhosteadyDoobper1}
\end{eqnarray}
so that it involves both corrections $r_0^{(1)}({\tilde x}) $ and $l_0^{(1)}({\tilde x}) $ of Eq. \ref{eigenvecper1}.

The second-order correction $\lambda^{(2)} $ also involves the Green function of Eq. \ref{green}
\begin{eqnarray}
 && \lambda^{(2)}  = \langle  l_0 \vert w^{(2)} \vert r_0   \rangle 
  + \langle  l_0 \vert w^{(1)} G w^{(1)}\vert r_0   \rangle 
  \nonumber \\
  && =    \int d {\tilde x} \int dx l_0 ({\tilde x}) w^{(2)}({\tilde x} \vert x)  r_0 (x)
  +  \int d {\tilde x}_2 \int dx_2 \int d {\tilde x}_1 \int dx_1 l_0 ({\tilde x}_2) w^{(1)}({\tilde x}_2,x_2)  
    w^{(1)} ({\tilde x}_1 \vert x_1)  r_0 (x_1)
   \nonumber \\
  && =    \int d {\tilde x} \int dx  w({\tilde x} \vert x) \frac{ O^2 ( {\tilde x},x)} {2}  \rho_* (x)
    +  \int d {\tilde x}_2 \int dx_2 \int d {\tilde x}_1 \int dx_1  w({\tilde x}_2,x_2)   O ( {\tilde x}_2,x_2)
    W ({\tilde x}_1 \vert x_1)  O ( {\tilde x}_1,x_1) \rho_* (x_1)
    \ \ \ 
\label{lambdaper2}
\end{eqnarray}


\subsection{ Large deviations at higher levels }

\label{subsec_higherlevels}

\subsubsection{ Explicit large deviations at level 2.5 for the empirical density $ {\mathring  \rho}(.)$ 
and the empirical kernel $  {\mathring w}(.,.) $ }

If one observes a trajectory $(x_0,x_1,...,x_T)$ over a long-time $T$,
it is interesting to consider 

(i) the empirical 1-point density
\begin{eqnarray}
{\mathring  \rho}( x)  \equiv \frac{1}{T} \sum_{t=0}^{T-1}  \delta (  x_t-  x)
\label{rho1pt}
\end{eqnarray}
satisfying the normalization
\begin{eqnarray}
\int d  x {\mathring  \rho} (x) = 1
\label{rho1ptnorma}
\end{eqnarray}

(ii) the empirical kernel 
\begin{eqnarray}
{\mathring w}( {\tilde x},  x) 
= \frac{ \displaystyle \frac{1}{T} \sum_{t=0}^{T-1} \delta (  x_{t+1}-  {\tilde x}) \delta (  x_t-  x)  }
{ \displaystyle \frac{1}{T} \sum_{t=0}^{T-1}  \delta (  x_t-  x) } 
= \frac{ \displaystyle \frac{1}{T} \sum_{t=0}^{T-1} \delta (  x_{t+1}-  {\tilde x}) \delta (  x_t-  x)  }{{\mathring  \rho}(  x)  } 
\label{empiw}
\end{eqnarray}
satisfying the normalization
\begin{eqnarray}
\int d{\tilde x} {\mathring w}( {\tilde x},  x) && \equiv \frac{ \displaystyle \frac{1}{T} \sum_{t=0}^{T-1}  \delta (  x_t-  x)  }{{\mathring  \rho} (  x)  } =1 \ \ \ \text{ for any }  x
\label{empiwnorma}
\end{eqnarray}
and having the empirical density ${\mathring  \rho} (.) $ of Eq. \ref{rho1pt} as steady distribution
\begin{eqnarray}
\int dx {\mathring w}( {\tilde x},  x) {\mathring  \rho} (x) =   \frac{1}{T} \sum_{t=0}^{T-1} \delta (  x_{t+1}-  {\tilde x}) 
= {\mathring  \rho} ({\tilde x}) + \frac{ \delta(x_T-{\tilde x}) - \delta(x_0-{\tilde x})}{T}   = {\mathring  \rho} ({\tilde x}) + O\left( \frac{1}{T} \right)
\label{empiwsteady}
\end{eqnarray}
up to boundary terms of order $1/T$ that are negligible for large $T$.

The joint distribution $P^{[2.5]}_T \left[ {\mathring  \rho} (.) ; {\mathring w}(.,.) \right]  $
of the empirical density ${\mathring  \rho} (.) $ of Eq. \ref{rho1pt}
and of the empirical kernel $ {\mathring w}(.,.) $ of Eq. \ref{empiw}
displays the large deviation form for large $T$ \cite{fortelle_thesis,fortelle_chain,review_touchette,c_largedevdisorder,c_reset,c_inference,c_microcanoEnsembles,c_diffReg}
\begin{eqnarray}
P^{[2.5]}_T \left[ {\mathring  \rho} (.) ; {\mathring w}(.,.) \right] 
&& \opsimeq_{T \to + \infty}
   \delta \left( \int d  x {\mathring  \rho}( x)  - 1 \right) 
  \left[ \prod_{ x} \delta \left(   \int d{\tilde x} {\mathring w}( {\tilde x},x) - 1 \right) \right]
 \left[\prod_{ {\tilde x}} \delta \left(   \int d  x   {\mathring w}( {\tilde x},  x) {\mathring  \rho} (x)- {\mathring  \rho} ({\tilde x}) \right)  \right] \ \nonumber \\
 && e^{\displaystyle -T  I_{2.5}\left[ {\mathring  \rho} (.) ; {\mathring w}(.,.) \right] }
\label{pnormaempi}
\end{eqnarray}
where the factors in front of the exponential correspond to the constitutive constraints 
for the empirical observables discussed in Eqs \ref{rho1ptnorma} \ref{empiwnorma} \ref{empiwsteady},
while the rate function $I_{2.5}\left[ {\mathring  \rho} (.) ; {\mathring w}(.,.) \right] $ at level 2.5 
appearing in the exponential has for explicit expression
\begin{eqnarray}
  I_{2.5} [ {\mathring  \rho} (.) ; {\mathring w}(.,.) ]
  =   
 \int d  x {\mathring  \rho}(x) \int d{\tilde x} {\mathring w}( {\tilde x},  x) \ln \left( \frac{{\mathring w}( {\tilde x},  x)}{ w( {\tilde x},  x)  }  \right) 
\label{rate2.5}
\end{eqnarray}

This level 2.5 for large deviations 
plays an essential role since it is the lowest level that is explicit
for an arbitrary Markov chain. Indeed, all the lower levels can be obtained via optimization
of this level 2.5 in the presence of constraints, but the solution of these
constrained optimizations cannot be written explicitly in general, as recalled below.


\subsubsection{ Large deviations at level 2 for the empirical density $ {\mathring  \rho}(.)$ alone }

The level 2 concerning the distribution
of the empirical density $ {\mathring  \rho}(.)$ alone
can be obtained from the integration over Eq. \ref{pnormaempi} 
over the empirical kernel ${\mathring w}(.,.)$
\begin{eqnarray}
&& P^{[2]}_T \left[  {\mathring  \rho}(.)  \right] \equiv  \int {\cal D} {\mathring w}(.,.)(.,.)  P^{[2.5]}_T \left[ {\mathring  \rho} (.) ; {\mathring w}(.,.) \right] 
 \nonumber \\
 && \opsimeq_{T \to + \infty} 
  \delta \left( \int d  x {\mathring  \rho}( x)  - 1 \right) 
 \int {\cal D} {\mathring w}(.,.)(.,.)  
 \left[ \prod_{ x} \delta \left(   \int d{\tilde x} {\mathring w}( {\tilde x},x) - 1 \right) \right]
 \left[\prod_{ y} \delta \left(   \int d  x   {\mathring w}( {\tilde x},  x) {\mathring  \rho} (x)- {\mathring  \rho} ({\tilde x}) \right)  \right] \  e^{\displaystyle -T  I_{2.5}\left[ {\mathring  \rho} (.) ; {\mathring w}(.,.) \right] }
\nonumber \\
  &&
 \opsimeq_{T \to + \infty}
 \delta \left( \int d  x {\mathring  \rho}( x)  - 1 \right)   e^{\displaystyle -T  I_{2}\left[ {\mathring  \rho}(.)   \right] }
\label{p2rho}
\end{eqnarray}
So the rate function $I_{2}\left[  {\mathring  \rho}(.)  \right] $ at level 2 for the empirical density alone
corresponds to the optimization of the rate function $I_{2.5}\left[ {\mathring  \rho} (.) ; {\mathring w}(.,.) \right] $ at level 2.5 over the empirical kernel ${\mathring w}(.,.)$ satisfying the appropriate constraints,
but cannot be written as an explicit function of ${\mathring  \rho}(.)  $ in general,
since the solution of this constrained optimization is not explicit for an arbitrary Markov chain.

\subsubsection{ Link with the large deviations of trajectory observables    }

The trajectory observable ${\cal O}^{Traj} [ x_0,x_1,...,x_T ]$ 
of Eq. \ref{observablerho2}
can be rewritten as a function $ {\cal O} 
 \left[ {\mathring  \rho} (.) ; {\mathring w}(.,.) \right]  $
 of the empirical density ${\mathring  \rho} (.) $ of Eq. \ref{rho1pt}
and of the empirical kernel $ {\mathring w}(.,.) $ of Eq. \ref{empiw}
\begin{eqnarray}
{\cal O}^{Traj}\left[ x_0,x_1,...,x_T \right]  
 =   \int d{\tilde x}  \int d  x \Omega( {\tilde x} ,  x)  {\mathring w}({\tilde x},x) {\mathring \rho}( x ) 
\equiv  {\cal O} \left[ {\mathring  \rho} (.) ; {\mathring w}(.,.) \right]
\label{observablerho2empi}
\end{eqnarray}

As a consequence, 
the probability distribution $P_T({\cal O}) $ of the trajectory observable $ $
already discussed in Eq. \ref{ldtraj}
 can be alternatively evaluated from 
  the probability $P^{[2.5]}_T \left[ {\mathring  \rho} (.) ; {\mathring w}(.,.) \right]   $ at the Level 2.5 of Eq. \ref{pnormaempi} via the integral
\begin{eqnarray}
P_T({\cal O})  && =\int {\cal D}   {\mathring \rho}(.,.)(.)  
\int {\cal D}    {\mathring w}(.,.)(.,.)  
P^{[2.5]}_T \left[ {\mathring  \rho} (.) ; {\mathring w}(.,.) \right] 
\delta \left({\cal O} - {\cal O} \left[{\mathring  \rho} (.) ; {\mathring w}(.,.)  \right] \right)
\nonumber \\
&& \opsimeq_{T \to +\infty} 
\int {\cal D}   {\mathring \rho}(.,.)(.)  
\int {\cal D}    {\mathring w}(.,.)(.,.)  
   \delta \left( \int d  x {\mathring  \rho}( x)  - 1 \right) 
  \left[ \prod_{ x} \delta \left(   \int d{\tilde x} {\mathring w}( {\tilde x},x) - 1 \right) \right]
 \left[\prod_{ y} \delta \left(   \int d  x   {\mathring w}( {\tilde x},  x) {\mathring  \rho} (x)- {\mathring  \rho} ({\tilde x}) \right)  \right] \ 
\nonumber \\
 &&  \delta \left({\cal O} - {\cal O} \left[{\mathring  \rho} (.) ; {\mathring w}(.,.)  \right] \right) e^{\displaystyle -T  I_{2.5}\left[ {\mathring  \rho} (.) ; {\mathring w}(.,.) \right] }
 \opsimeq_{T \to +\infty} e^{- T I({\cal O})}
\label{largedevadditiveproba}
\end{eqnarray}
The daddle-point evaluation of the integral for large $T$
means that the rate function $I({\cal O})$ for the trajectory observable ${\cal O} $
already discussed around Eq. \ref{legendre} 
also corresponds to the optimization of the explicit rate function $I^{[2.5]}\left[ {\mathring  \rho} (.) ; {\mathring w}(.,.) \right]  $
over the empirical density ${\mathring  \rho} (.) $ 
and the empirical kernel $ {\mathring w}(.,.) $ satisfying their constitutive constraints,
as well as the supplementary constraint $ \delta \left({\cal O} - {\cal O} \left[{\mathring  \rho} (.) ; {\mathring w}(.,.)  \right] \right)$ reproducing the appropriate value ${\cal O}$ of the trajectory observable.
Again, the solution of this constrained optimization is not explicit for an 
arbitrary observable of an arbitrary Markov chain, so the rate function $I({\cal O}) $ 
can be computed explicitly only in special cases.


\section{ Alternative analysis of the large deviations and conditioning via the backward Markov chain}

\label{app_backward}

In this Appendix, we describe how the large deviations and conditioning 
analyzed via the forward Markov chain in the previous Appendix \ref{app_markov}
can be alternatively studied from the point of view of the backward Markov chain,
since this change of perspective is very useful in the main text concerning chaotic non-invertible maps.

\subsection{ Backward Markov chain associated to the forward Markov chain}

When the initial condition $x_0$ is drawn with the steady state distribution $\rho_*(x_0)$,
the probability ${\cal P}_*^{Forward} (x_{0 \leq t \leq T} ) $ of the forward trajectory $(x_0,x_1,..,x_T )$
over the time-window $[0,T]$ reads in terms of the forward-kernel $w(. \vert .) $ 
\begin{eqnarray}
{\cal P}_T^{Traj} (x_0,x_1,..,x_T )  =   \rho_*(x_0 )  \prod_{t=0}^{T-1}  w( x_{t+1} \vert x_t \big ) 
\label{Ptrajforward}
\end{eqnarray}

For $T=1$, the joint probability of the two consecutive positions $(x_0,x_1)$ can be also rewritten as
\begin{eqnarray}
{\cal P}_{T=1}^{Traj} (x_0,x_1 ) && = w( x_1 \vert x_0) \rho_*(x_0 ) 
 \equiv   w_B(x_0 \vert x_1) \rho_*(x_1)
\label{Ptrajforward2points}
\end{eqnarray}
where the backward-kernel $w^B(x \vert {\tilde x}) $ 
represents the probability to see the previous point $x$ when the next point is $\tilde x$ 
and can be obtained from the forward kernel $w( {\tilde x} \vert x) $ via the similarity transformation involving the steady state $\rho_*$ 
\begin{eqnarray}
w_B(x \vert {\tilde x})  = \frac{1}{\rho_*({\tilde x})} w( {\tilde x} \vert x) \rho_*(x ) 
\label{Wbackward}
\end{eqnarray}
The spectral decomposition of Eq. \ref{Wspectral} of the forward kernel $w( {\tilde x} \vert  x) $
yields that the spectral decomposition of the backward kernel
\begin{eqnarray}
w_B(x \vert {\tilde x})   
= \sum_{p=0}^{+\infty} \lambda_p  \left[ \rho_*(x ) l_p(x) \right]\left[ \frac{r_p({\tilde x})}{ \rho_*({\tilde x})} \right]
\equiv \sum_{p=0}^{+\infty} \lambda_p r^B_p(  x)l^B_p(  {\tilde x})
\label{Wbackwardspectral}
\end{eqnarray}
involves the same eigenvalues $\lambda_p$ as the forward kernel $w( {\tilde x} \vert x)  $,
while the right and left eigenvectors for the backward kernel read
\begin{eqnarray}
r^B_p(  x) && =l_p(  x)  \rho_*(x)
\nonumber \\
l^B_p(  {\tilde x}) &&  = \frac{ r_p(  {\tilde x}) }{ \rho_*({\tilde x})} 
\label{eigenvecWB}
\end{eqnarray}

In particular, for the highest eigenvalue $\lambda_{p=0}=1$, 
plugging Eq. \ref{eigenW1} into Eq. \ref{eigenvecWB} yields
\begin{eqnarray}
r^B_0(  {\tilde x}) && =  \rho_*({\tilde x})
\nonumber \\
l^B_0(  x) &&  = 1
\label{eigenvecWBzero}
\end{eqnarray}
This means that
the backward-kernel of Eq. \ref{Wbackward} is normalized over $x$ for any $\tilde x$
and that
the backward kernel of Eq. \ref{Wbackward} has the same steady state $\rho_*(.)$ as the forward kernel $w(. \vert .)$
\begin{eqnarray}
 \rho_*(x )  = \int d {\tilde x} w_B(x \vert {\tilde x}) \rho_*({\tilde x})  
\label{Wbackwardsteady}
\end{eqnarray}

The backward-kernel $w^B(x \vert {\tilde x}) $
can be used to construct the trajectory going backward 
since the trajectory probability of Eq. \ref{Ptrajforward}
can be rewritten as
\begin{eqnarray}
{\cal P}_T^{Traj} (x_0,x_1,..,x_T ) && =   \rho_*(x_0 )  \prod_{t=0}^{T-1}  w( x_{t+1} \vert x_t \big ) 
= \rho_*(x_0 )  
\prod_{t=0}^{T-1} \left[ w_B( x_t \vert x_{t+1}  \big ) \frac{\rho_*(x_{t+1})} {\rho_*(x_t)}\right]
\nonumber \\
&& =  \rho_*(x_T ) \prod_{t=0}^{T-1}  w_B( x_t \vert x_{t+1}  \big )
\label{Ptrajforwardback}
\end{eqnarray}
When considering the backward trajectory, it is thus useful to replace the forward-time $t=0,1,..,T$ by
the backward-time  
\begin{eqnarray}
 \tau= T-t  =0,1,..,T
\label{backwardtime}
\end{eqnarray}
and to relabel the positions $x_t=y_{(\tau=T-t)}$ :
then the backward time $\tau=0,1,..,T$ grows  
from the minimal value $\tau=0$ associated the initial position $x_T=y_0$ of the backward trajectory 
towards the maximal value $\tau=T$ associated the final position $x_0=y_T$ of the backward trajectory.


\subsection{ Large deviations of trajectory observables via the deformed backward dynamics }

\subsubsection{  Generating function via the $k$-deformed backward kernel $ w_B^{[k]}(x \vert {\tilde x})$  }

The $k$-deformed kernel of Eq. \ref{Wkdeformed} associated to the backward kernel $w_B(x \vert {\tilde x}) $ of Eq. \ref{Wbackward}
\begin{eqnarray}
 w_B^{[k]}(x \vert {\tilde x}) && \equiv w_B(x \vert {\tilde x}) e^{ k O ({\tilde x},x)}
 =   w({\tilde x} \vert x) \frac{ \rho_*(x ) }{ \rho_*({\tilde x})} e^{ k O ({\tilde x},x)}
 =\frac{1}{\rho_*({\tilde x})} w^{[k]}({\tilde x} \vert x)  \rho_*(x ) 
\label{Wbackwardkdeformed}
\end{eqnarray}
This similarity transformation between the deformed forward kernel $w^{[k]}({\tilde x} \vert x) $
and the deformed backward kernel $w_B^{[k]}(x \vert {\tilde x}) $
yields that their highest eigenvalue $ \lambda_0^{[k]} $ is the same,
while the positive right and left eigenvectors of the backward kernel
satisfying
\begin{eqnarray}
\lambda_0^{[k]} \ r_0^{B[k]}(  x) && =  \int d {\tilde x}  \ w^{[k]}_B(x \vert {\tilde x})  r_0^{B[k]}(  {\tilde x})  
\nonumber \\
\lambda_0^{[k]}  l_0^{B[k]}(  {\tilde x}) &&  =  \int d  x  l_0^{B[k]}(  x) w_B^{[k]}(x \vert {\tilde x})  
\label{eigenWkback}
\end{eqnarray}
are related to the forward eigenvectors via
\begin{eqnarray}
r_0^{B[k]}(  {\tilde x}) && =l_0^{[k]}(  {\tilde x})  \rho_*({\tilde x})
\nonumber \\
l_0^{B[k]}(  x) &&  = \frac{ r_0^{[k]}(  x) }{ \rho_*(x)} 
\label{eigenvecWBk}
\end{eqnarray}


\subsubsection{Construction of the backward conditioned steady state $\rho^{BC[k]}_*( x)  $
and of the backward conditioned kernel $w^{BC[k]}({\tilde x} \vert x) $  }

Using Eq. \ref{eigenvecWBk}, one obtains that 
the backward conditioned steady state $\rho^{BC[k]}_*( x)  $ coincides with the
forward conditioned steady state $\rho^{C[k]}_*( x)  $ of Eq. \ref{rhosteadyDoob}
\begin{eqnarray}
\rho^{BC[k]}_*( x) \equiv l_0^{B[k]}(  x) r_0^{B[k]}(  x)
=  l_0^{[k]}(x) r_0^{[k]}(x) \equiv \rho^{C[k]}_*( x) 
  \label{rhosteadyDoobbackward}
  \end{eqnarray}
  
The backward  conditioned kernel $w^{C[k]}_B(x \vert {\tilde x}) $  
constructed via the analog to Eq. \ref{WDoobk} 
\begin{eqnarray}
w^{C[k]}_B(x \vert {\tilde x}) 
&& \equiv  \frac{1}{\lambda_0^{[k]}} \  l_0^{B [k]}(x) w^{B[k]}(x \vert {\tilde x})   \frac{1}{ l_0^{B[k]}({\tilde x})  }
\nonumber \\
&& = \frac{l_0^{B[k]}(x) w^{B[k]}(x \vert {\tilde x})}{\int d  z  l_0^{B[k]}(  z) w^{B[k]}(z \vert {\tilde x})  }
= \frac{l_0^{B[k]}(x) w_B(x \vert {\tilde x}) e^{ k O ( {\tilde x},x)}}{\int d  z  l_0^{B[k]}(  z) w(z \vert {\tilde x}) e^{ k O ( {\tilde x},z)} }
  \label{WDoobkback}
\end{eqnarray}
is related to the conditioned forward kernel $w^{C[k]}({\tilde x} \vert x) $  
of Eq. \ref{WDoobk} via the
similarity transformation involving common conditioned steady density of Eq. \ref{rhosteadyDoobbackward}
\begin{eqnarray}
w^{C[k]}_B(x \vert {\tilde x}) \rho^{C[k]}_*( {\tilde x})  = w^{C[k]}({\tilde x} \vert x) \rho^{C[k]}_*( x)
  \label{WDoobkbackforw}
\end{eqnarray}
This equation 
represents the conditioned probability of two consecutive positions
as computed from the backward or forward perspectives,
and is the analog of Eq. \ref{Ptrajforward2points} for the initial dynamics.


\section { Doubling map : Properties of the forward and backward dynamics for the Fourier coefficients }

\label{app_fourier}

In order to better understand 
the forward and backward dynamics that were discussed for the density $\rho_.(.)$
on the real-space interval $x \in [0,1[$
in sections \ref{sec_forward} and \ref{sec_backward} of the main text,
it is useful in the present Appendix to describe
these dynamics from the point of view of the Fourier coefficients of the density $\rho_.(.)$
(see \cite{linas} for more details).


\subsection{ Forward dynamics for the Fourier coefficients ${\hat \rho_t}(n ) = \int_0^1  d {\tilde x}\rho_t(x ) e^{- i 2 \pi n x} $
of the density $\rho_t(  x)  $  }


\subsubsection{ General properties of the Fourier coefficients ${\hat \rho_t}(n ) = \int_0^1  d {\tilde x}\rho_t(x ) e^{- i 2 \pi n x} $ with $n \in \mathbf{Z}$ }

The Fourier series representation of the density $\rho_t(x ) $
\begin{eqnarray}
\rho_t(x ) =  \sum_{n=-\infty}^{+\infty} {\hat \rho_t}(n ) e^{i 2 \pi n x} 
\label{rhofourier}
\end{eqnarray}
involves the Fourier coefficients
\begin{eqnarray}
 {\hat \rho_t}(n ) = \int_0^1 dx \rho_t(x ) e^{- i 2 \pi n x} 
\label{rhofouriercoefs}
\end{eqnarray}
that satisfy the complex-conjugate relations (since the density is real
$\rho_t({\tilde x} ) = \overline{ \rho_t({\tilde x} )} $)
\begin{eqnarray}
 {\hat \rho}_t( -n ) = \int_0^1 dx \rho_t(x ) e^{ i 2 \pi n x}  = \overline{ {\hat \rho}_t( n )}
\label{rhofouriercoefscc}
\end{eqnarray}
while the coefficient for $n=0$ is fixed by the normalization
\begin{eqnarray}
 {\hat \rho}_t(n=0 ) = \int_0^1 dx \rho_t(x ) =1
\label{rhofouriercoefs0}
\end{eqnarray}


\subsubsection{ Frobenius-Perron dynamics for the Fourier coefficients ${\hat \rho_t}(n ) $ with $n \in \mathbf{Z}$}

The Frobenius-Perron evolution of Eq. \ref{doublingFP}
translates for the Fourier coefficients into
\begin{eqnarray}
 {\hat \rho}_t( {\tilde n} ) = \sum_{n=-\infty}^{+\infty} {\hat W}( {\tilde n} \vert n) {\hat \rho}_{t-1}(n ) 
\label{perronFFourier}
\end{eqnarray}
where the matrix ${\hat W}({\tilde n} \vert n) $ can be computed from the real-space kernel $w({\tilde x} \vert x) $ of Eq. \ref{Wdoubling}
\begin{eqnarray}
 {\hat W}({\tilde n} \vert n)  \equiv   \int_0^1 dx e^{i 2 \pi n x} \int_0^1  d {\tilde x}e^{-i 2 \pi {\tilde n} {\tilde x}}w({\tilde x} \vert x)
  =   \delta_{2{\tilde n},n}
\label{kernelFourier}
\end{eqnarray}
So the dynamics of Eq. \ref{perronFFourier} reduces to
\begin{eqnarray}
 {\hat \rho}_t( {\tilde n} ) = \sum_{n=-\infty}^{+\infty}  \delta_{2{\tilde n},n}  {\hat \rho}_{t-1}(n ) 
 = {\hat \rho}_{t-1}(2 {\tilde n})
\label{perronFFourierdoubling}
\end{eqnarray}
and corresponds to the Fourier translation of the real-space evolution of Eq. \ref{doublingFP} :
all odd Fourier coefficients ${\hat \rho}_{t-1}(n=2m+1) $ of the density at time $(t-1)$ disappear,
while the even Fourier coefficients ${\hat \rho}_{t-1}(2 {\tilde n}) $ become $ {\hat \rho}_t( {\tilde n} ) $.
The iteration of Eq. \ref{perronFFourierdoubling} up to the initial condition at $t=0$
\begin{eqnarray}
{\hat \rho}_t(  n)  = {\hat \rho}_{t-1} (2n) ={\hat \rho}_{t-2} (4n) 
 = ... = {\hat \rho}_0 (2^t n) 
\label{perronFFFourierdoublingiter}
\end{eqnarray}
shows that only the Fourier coefficients  ${\hat \rho}_0 (n_0)$ of the initial density 
with indices $n_0=2^t n$
have not yet disappeared at time $t$.
If the Fourier coefficients ${\hat \rho}_0 (n_0)  $ of the initial condition
decay to zero for large $n_0$, 
then
 only the Fourier coefficient $n=0$ survives for large time $t \to + \infty$
\begin{eqnarray}
{\hat \rho}_t(  n) \opsimeq_{t \to + \infty} {\hat \rho}_*(  n)  =\delta_{n=0}
\label{perronFrobsteadyfourier}
\end{eqnarray}
in correspondence with the uniform real-space steady state of Eq. \ref{perronFrobsteady}.

In Fourier space, the right eigenvectors ${\hat r}_p(  n)  $ associated to the eigenvalues $\lambda_p=2^{-p} $ of Eq. \ref{eigenvaluesdoubling}
satisfy
\begin{eqnarray}
\lambda_p \ {\hat r}_p(  n)   =   {\hat r}_p( 2 n)
\label{eigenWmapdoublingrfourier}
\end{eqnarray}
and are given by the Fourier coefficients of the Bernoulli polynomials $r_p(x)=B_p(x) $ of Eq. \ref{bernouilli}
\begin{eqnarray}
{\hat r}_p(  n) = {\hat B}_p(  n) = - \frac{ p! }{ (i 2 \pi n)^p} \ \ \text{ for } \ n \ne 0
\label{bernoullifourier}
\end{eqnarray}
with the first members
\begin{eqnarray}
{\hat r}_1(  n) = {\hat B}_1(  n) = - \frac{ 1 }{ i 2 \pi n} \ \ \text{ for } \ n \ne 0
\nonumber \\
{\hat r}_2(  n) = {\hat B}_2(  n) =  \frac{ 1 }{ 2 \pi^2 n^2} \ \ \text{ for } \ n \ne 0
\label{bernoullifourierfirst}
\end{eqnarray}

However, if one tries to translate the real-space left eigenvectors $ l_p(  x)  = D_p(x) $ of Eq. \ref{lndeltasingular} for $p=1,2,..$ in the Fourier basis, one obtains that all coefficients vanish.
And if one writes the left eigenvalue equation directly in terms of Fourier coefficients
\begin{eqnarray}
\lambda_p \ {\hat l}_p(  m)  =  \sum_{n=-\infty}^{+\infty} {\hat l}_p(  n)\delta_{2n,m}  
   = \begin{cases}
0 \text{  for odd $m$}  
 \\
{\hat l}^{[p]}( \frac{m}{2}) \text{ for even $m$ }
\end{cases}
\label{eigenWmapdoublingrfourierleft}
\end{eqnarray}
one cannot obtain any solution for $p \ne 0$.



\subsection{ Backward dynamics for the Fourier coefficients }

In Fourier space with Eqs \ref{rhofourier} and \ref{rhofouriercoefs}, 
the backward dynamics of Eq. \ref{doublingFPbackward}
translates for the Fourier coefficients into
\begin{eqnarray}
 {\hat \rho}^B_{\tau}( n ) = \sum_{{\tilde n}=-\infty}^{+\infty} {\hat W}_B(n \vert {\tilde n}) {\hat \rho}^B_{\tau-1}({\tilde n} ) 
\label{perronFFourierbackdef}
\end{eqnarray}
with the matrix
\begin{eqnarray}
 {\hat W}^B(n \vert {\tilde n})  
 \equiv   \int_0^1 dx e^{i 2 \pi n x} \int_0^1  d {\tilde x}e^{-i 2 \pi {\tilde n} {\tilde x}}w({\tilde x} \vert x)
  =   \delta_{n,2 {\tilde n}} 
\label{kernelFourierback}
\end{eqnarray}
so that Eq. \ref{perronFFourierbackdef}
reduces to
\begin{eqnarray}
 {\hat \rho}^B_{\tau}( n ) = \sum_{{\tilde n}=-\infty}^{+\infty}  \delta_{n,2 {\tilde n}}  {\hat \rho}^B_{\tau-1}({\tilde n} ) 
    = \begin{cases}
{\hat \rho}^B_{\tau-1} ( \frac{n}{2}) \text{ for even $n$ }
 \\
0 \text{  for odd $n$}  
\end{cases}
\label{perronFFourierback}
\end{eqnarray}
The iteration up to the initial condition at $\tau=0$ yields
\begin{eqnarray}
 {\hat \rho}^B_{\tau}( n )     = \begin{cases}
 {\hat \rho}^B_0 ( \frac{n}{2^{\tau}}) \text{ if $\frac{n}{2^{\tau}}=n_0$ is an integer}
 \\
0 \text{  if $\frac{n}{2^{\tau}}$ is not an integer}  
\end{cases}
\label{perronFFourierbackiter}
\end{eqnarray}
The physical meaning is that the Fourier coefficients 
${\hat \rho}^B_0 ( n_0) $ of the initial condition at $\tau=0$
are transferred to the Fourier coefficients $ {\hat \rho}^B_{\tau}( n=2^{\tau} n_0 ) $
while all the other coefficients where $ \frac{n}{2^{\tau}}$ is not an integer are zero.

For large time $\tau \to + \infty$,
 only the Fourier coefficient $n=0$ survives 
\begin{eqnarray}
{\hat \rho}^B_{\tau}( n ) \opsimeq_{\tau \to + \infty} {\hat \rho}^B_*(  n)  =\delta_{n,0}
\label{perronFrobsteadyfourierback}
\end{eqnarray}
in correspondence with the uniform real-space steady state of Eq. \ref{perronFrobsteadyback}.
However this convergence is very weird, since for any finite large $\tau$,
there are finite isolated Fourier coefficients for $n=2^{\tau} n_0 $ in Eq. \ref{perronFFourierback}.

Via the exchange of Eq. \ref{eigenvecWB}
between the right and left eigenvectors between the forward and backward kernels,  one obtains
that the left eigenvectors of the backward Fourier kernel ${\hat W}^B(n \vert {\tilde n})   $ of Eq. \ref{kernelFourierback}
are given by Eq. \ref{bernoullifourier}
\begin{eqnarray}
{\hat l}^B_p(  n) = {\hat r}_p(  n) = {\hat B}_p(  n) = - \frac{ p! }{ (i 2 \pi n)^p} \ \ \text{ for } \ n \ne 0
\label{bernoullifourierB}
\end{eqnarray}
and correspond to the Fourier coefficients of the Bernoulli polynomials $B_p(x) $
of Eq. \ref{bernouilli},
while there are no right eigenvectors for $p \ne 0$,
since there are no left eigenvectors for the Fourier forward kernel
as discussed around Eq. \ref{eigenWmapdoublingrfourierleft}.


\end{document}